\newif\ifsubmode
\submodefalse

\ifsubmode
\documentclass[10pt,preprint]{aastex}
  \received{}
  \revised{}
  \accepted{}
\else
  \documentclass[iop]{emulateapj}   \usepackage{apjfonts}
\fi

\usepackage{natbib}
\usepackage{epsf}
\usepackage{color}
\usepackage{amsmath}
\ifsubmode
  
\else
  
\fi

\newcommand{\hst}{\textit{HST}}

\newcommand{\spitzer}{\textit{Spitzer}}

\newcommand{\chandra}{\textit{Chandra}}

\newcommand{\nich}{\hbox{$H_{160}$}}

\newcommand{\wfcj}{\hbox{$J_{125}$}}
\newcommand{\wfch}{\hbox{$H_{160}$}}

\newcommand{\lsim}{\lesssim}
\newcommand{\gsim}{\gtrsim}

\newcommand{\eg}{e.g.}

\newcommand{\msol}{\hbox{$M_\odot$}}

\newcommand{\uJy}{\hbox{$\mu$Jy}}
\newcommand{\ujy}{\hbox{$\mu$Jy}}

\newcommand{\mone}{\hbox{$[3.6]$}}

\newcommand{\fulltarget}{XMM--LSS~J02182-05102}

\newcommand{\bs}{\hbox{$\!\!\!\!$}}
\newcommand{\sersic}{S\'ersic}
\ifsubmode
\else
\renewcommand{\bs}{\hbox{}}
\fi

\shorttitle{GALAXY STRUCTURE IN A CLUSTER AT $z=1.62$}
\shortauthors{PAPOVICH ET AL.}

\begin{document}

\slugcomment{\it Submitted for publication in the Astrophysical Journal}
\ifsubmode
\title{CANDELS OBSERVATIONS OF THE STRUCTURAL PROPERTIES AND EVOLUTION OF GALAXIES IN A CLUSTER AT $Z$=1.62}
\else
\title{CANDELS OBSERVATIONS OF THE STRUCTURAL PROPERTIES OF CLUSTER GALAXIES AT $Z$=1.62}
\fi

\author{\sc C.~Papovich\altaffilmark{1}, 
   R.~Bassett\altaffilmark{1},
  J.~M.~Lotz\altaffilmark{2}, 
A.~van der Wel\altaffilmark{3}, 
K.-V~Tran\altaffilmark{1}, 
S.~L.~Finkelstein\altaffilmark{1,4,5}, 
E.~F.~Bell\altaffilmark{6},
C.~J.~Conselice\altaffilmark{7},
A.~Dekel\altaffilmark{8},
J.~S.~Dunlop\altaffilmark{9}, 
Yicheng~Guo\altaffilmark{10},
S.~M.~Faber\altaffilmark{11},
D.~Farrah\altaffilmark{12},
H.~C.~Ferguson\altaffilmark{2},
K.~D.~Finkelstein\altaffilmark{1,4}, 
B.~H\"au\ss ler\altaffilmark{7}, 
D.~D.~Kocevski\altaffilmark{11},
A.~Koekemoer\altaffilmark{2}, 
D.~C.~Koo\altaffilmark{11},
E.~J.~McGrath\altaffilmark{11},
R.~J.~McLure\altaffilmark{9},
D.~H.~McIntosh\altaffilmark{13},
I.~Momcheva\altaffilmark{14}, 
J.~A.~Newman\altaffilmark{15},
G.~Rudnick\altaffilmark{16},
B.~Weiner\altaffilmark{17}, 
C.~N.~A.~Willmer\altaffilmark{17},
S.~Wuyts\altaffilmark{18}}
\ifsubmode
\altaffiltext{1}{George P.\ and Cynthia Woods Mitchell Institute for Fundamental Physics and Astronomy, and 
Department of Physics and Astronomy, Texas A\&M University, College Station, TX, 77843-4242; papovich@physics.tamu.edu}
\altaffiltext{2}{Space Telescope Science Institute, 3700 San Martin
  Dr., Baltimore, MD 21218}
\altaffiltext{3}{Max-Planck-Institut f\"ur Astronomie, K\"onigstuhl 17, D-69117, Heidelberg, Germany} 
\altaffiltext{4}{Department of Astronomy, University of Texas,
  Austin, TX 78712}
\altaffiltext{5}{Hubble Fellow}
\altaffiltext{6}{Department of Astronomy, University of Michigan,  Ann Arbor, MI 48109}
\altaffiltext{7}{School of Physics \& Astronomy, University of
  Nottingham, Nottingham, UK}
\altaffiltext{8}{Racah Institute of Physics, The Hebrew University, Jerusalem 91904, Israel}
\altaffiltext{9}{Institute for Astronomy, University of Edinburgh,
  Royal Observatory, Edinburgh, UK}
\altaffiltext{10}{Astronomy Department, University of Massachusetts, Amherst, MA 01003}
\altaffiltext{11}{UCO/ Lick Observatory, Department of Astronomy and Astrophysics, University of California, Santa Cruz, CA 95064}
\altaffiltext{12}{Astronomy Centre, University of Sussex, Falmer,
  Brighton, UK}
\altaffiltext{13}{Department of Physics, University of Missouri-Kansas City, 5110 Rockhill Road, Kansas City, MO 64110}
\altaffiltext{14}{Observatories Carnegie Institution of Washington, Pasadena, CA, 91101}
\altaffiltext{15}{University of Pittsburgh, Department of Physics and
Astronomy, Pittsburgh, PA 15260}
\altaffiltext{16}{Department of Physics and Astronomy, University of Kansas, Lawrence, KS, 66045-7582}
\altaffiltext{17}{Steward Observatory, University of Arizona, Tucson, AZ 85721}
\altaffiltext{18}{Max-Planck-Institut f\"{u}r extraterrestrische
  Physik, Giessenbachstrasse, D-85748 Garching, Germany}
\else
\affil{$^{1}$George P.\ and Cynthia Woods Mitchell Institute for Fundamental Physics and Astronomy, and 
Department of Physics and Astronomy, Texas A\&M University, College Station, TX, 77843-4242; papovich@physics.tamu.edu}
\affil{$^{2}$Space Telescope Science Institute, 3700 San Martin
  Dr., Baltimore, MD 21218}
\affil{$^{3}$Max-Planck-Institut f\"ur Astronomie, K\"onigstuhl 17, D-69117, Heidelberg, Germany} 
\affil{$^{4}$Department of Astronomy, University of Texas,
  Austin, TX 78712}
\affil{$^{5}$Hubble Fellow}
\affil{$^{6}$Department of Astronomy, University of Michigan,  Ann Arbor, MI 48109}
\affil{$^{7}$School of Physics \& Astronomy, University of
  Nottingham, Nottingham, UK}
\affil{$^{8}$Racah Institute of Physics, The Hebrew University, Jerusalem 91904, Israel}
\affil{$^{9}$Institute for Astronomy, University of Edinburgh,
  Royal Observatory, Edinburgh, UK}
\affil{$^{10}$Astronomy Department, University of Massachusetts, Amherst, MA 01003}
\affil{$^{11}$UCO/ Lick Observatory, Department of Astronomy and Astrophysics, University of California, Santa Cruz, CA 95064}
\affil{$^{12}$Astronomy Centre, University of Sussex, Falmer,
  Brighton, UK}
\affil{$^{13}$Department of Physics, University of Missouri-Kansas City, 5110 Rockhill Road, Kansas City, MO 64110}
\affil{$^{14}$Observatories Carnegie Institution of Washington, Pasadena, CA, 91101}
\affil{$^{15}$University of Pittsburgh, Department of Physics and
Astronomy, Pittsburgh, PA 15260}
\affil{$^{16}$Department of Physics and Astronomy, University of Kansas, Lawrence, KS, 66045-7582}
\affil{$^{17}$Steward Observatory, University of Arizona, Tucson, AZ 85721}
\affil{$^{18}$Max-Planck-Institut f\"{u}r extraterrestrische
  Physik, Giessenbachstrasse, D-85748 Garching, Germany}
\fi



\begin{abstract}  

\noindent 
We discuss the structural and morphological properties of galaxies in
a $z=1.62$ proto-cluster using near--IR imaging data from
\textit{Hubble Space Telescope} Wide Field Camera 3 data of 
the Cosmic Assembly Near-IR Deep Extragalactic Legacy Survey
(CANDELS).  The cluster galaxies exhibit a clear color--morphology
relation:  galaxies with colors of quiescent stellar populations
generally have morphologies consistent with spheroids, and galaxies
with colors consistent with ongoing star formation have disk--like and
irregular morphologies.  The size distribution of the quiescent
cluster galaxies shows a deficit of compact ($\lsim 1$~kpc), massive
galaxies compared to CANDELS field galaxies at $z=1.6$.  As a result the cluster quiescent galaxies
have larger average effective sizes compared to field galaxies at
fixed mass at greater than 90\% significance.   Combined with data
from the literature,  the size evolution of quiescent cluster galaxies
is relatively slow from $z\simeq 1.6$ to the present, growing as
$(1+z)^{-0.6\pm0.1}$.  If this result is generalizable, then it
implies that  physical processes associated with the denser cluster
region seems to have caused accelerated size growth in quiescent
galaxies prior to $z=1.6$ and slower subsequent growth at $z<1.6$
compared to galaxies in the lower density field.    The quiescent
cluster galaxies at $z=1.6$ have higher ellipticities compared to
lower redshift samples at fixed mass, and their surface-brightness
profiles suggest that they contain extended stellar disks.
We argue the cluster galaxies require dissipationless (i.e., gas--poor
or ``dry'') mergers to reorganize the disk material and to match the
relations for ellipticity, stellar mass, size, and color of early-type galaxies in $z<1$ clusters. 
\end{abstract}
 


\section{INTRODUCTION}

Massive elliptical and early--type galaxies dominate regions of high
density such as those of galaxy clusters in the present Universe
\citep[\eg,][]{dres80,post84}.  By $z\lsim 1.5$, \textit{Hubble Space
Telescope} (\hst)  observations show that these passive cluster
galaxies have elliptical and lenticular morphologies, with a strong
color-density relationship
\citep[\eg,][]{vandokkum01,blak03,post05,mei06,blak06,hilton09}.
The emerging picture for formation and evolution of the massive, red,
early-type cluster galaxies is one in which these galaxies formed
their stars at $z\gsim 2$, with subsequent passive evolution
\citep[\eg,][]{stan98,eise08,whiley08}.   These galaxies continue to
grow by mergers and secular processes, with negligible additional
star-formation in order for their color evolution to be consistent
with observations.

The details of this evolution is unclear, yet these cluster galaxies
must assemble \textit{sometime}.    It may be that the formation of
cluster galaxies is related to the cluster assembly process itself
\citep[\eg,][]{dubi98}.   Observations show that the intracluster
galaxy velocity dispersion is lower in forming clusters and groups,
and  therefore galaxy--galaxy interactions are more frequent
\citep[see,][]{vandokkum99,lidman08,mcintosh08,tran08,mcgee09,wilman09}.
Therefore, one may expect strong morphological evolution as a result
of increased mergers, which lead to a population of spherical,
elliptical galaxies \citep{nava90}.   Out to $z\sim 1$, ellipticals
dominate the galaxy populations of massive clusters
\citep{desai07,holden09,vulc11}, while lenticular and early-type
spiral galaxies dominate the cores of some lower density
groups \citep{wilman09,just10}.  If lower mass groups
are common precursors to galaxy clusters, then their galaxies must
undergo morphological evolution to early-type galaxies as the groups
merge to form larger clusters.   This is expected based on some
semianalytic models, which predict that processes associated with the
cluster formation are expected to influence galaxy evolution at $z\gg
1$ \citep{dubi98,lin04a,delucia07b,rusz09} where the main progenitors
of clusters collapse \citep[\eg,][]{boylan09}.  Therefore, if this
hypothesis is correct, then as we encroach on the formation epochs of
today's massive clusters, $z \gsim 1.5$, we should expect to see rapid
evolution in the properties of the cluster galaxies.    

In addition, observations show quiescent galaxies (not only those in
clusters), with apparent early-type morphologies,  undergo strong size
evolution with redshift out to $z=2$
\citep[\eg,][]{daddi05b,papo05,truj06,truj07,long07,zirm07,toft07,buit08,cima08,vanderwel08,vandokkum08a,damj09,cass10,sara10}.
One explanation for this size evolution is that these galaxies grow by
frequent  dissipationless (i.e., gas--poor or ``dry'') minor mergers
\citep[\eg,][]{loeb03,vandokkum05a,bell06,naab06a,naab06b,naab07,khoc06a,khoc06b,lotz08,masj08,hopk09b,hopk10b,vanderwel09,vanderwel11}
Minor mergers would cause the galaxies to add mass at larger radii,
increasing their effective sizes substantially with a relatively small
increase in stellar mass \citep{oser10}.  Some recent observations
support this interpretation \citep{beza09,hopk09,vandokkum10},
although this explanation would not explain the substantially larger
central densities of high--redshift ellipticals compared to galaxies
at lower redshift \citep[see \eg,][]{stoc10}.  Alternatively,
\citet{graham11} notes that many of these compact objects share sizes,
masses, and mass densities of present-day bulges, suggesting some of
these objects are the precursors to the spheroidal components of
present-day disk galaxies. 

It is unclear how the assembly of ellipticals in high density cluster
(and forming cluster) regions differs from that in the lower density
the field.    If the size growth of ellipticals is driven by minor
mergers and galaxies experience more mergers in forming clusters, then
it follows that the size and morphological evolution of cluster
ellipticals should be accelerated during the cluster formation stage.
\citet{stott11} report that the sizes of the most massive galaxies in
clusters increase by at most 30\% during the period of $z=1$ to 0.2.
\citet{cooper11} find a correlation between the sizes and local galaxy
overdensity for early-type galaxies at $0.4 < z < 1.2$, suggesting
accelerated morphological evolution in regions in higher density.
\citet{zirm11} find a hint of evidence that massive quiescent galaxies
in the vicinity radio galaxy MRC 1138$-$262 at $z=2.2$ have larger
sizes at fixed mass compared to galaxies in the field at this
redshift.  These observations support the hypothesis that cluster
ellipticals experience accelerated structural evolution.  However,
other observations at higher redshift ($z\sim 2.3-4.1$) find no
evidence that the sizes or morphologies of galaxies differ in
high--density regions compared to those of low--density regions
\citep[\eg][]{peter07,over08}, suggesting that any environmental
effects are not yet present at these epochs.

Here, we compare the properties of galaxies in the high density region
of a forming cluster at $z=1.62$, \fulltarget\
\citep{papo10a,tanaka10}, and we compare them to similarly selected
galaxies in the lower density $z=1.6$ field.  This galaxy cluster was
identified as an overdensity of sources with \spitzer/IRAC colors
indicative of high--redshift galaxies \citep{papo08}.   The cluster
shows a dominant population of red galaxies, which form a strong ``red
sequence'' population, with an estimate of the last major
star--formation epoch of $z_f = 2.2-2.3$ \citep{papo10a}.  In
addition, this cluster shows a significant fraction of star--forming
galaxies as evidenced by their \spitzer/24~\micron\ emission
\citep{tran10}.    There are currently 13 redshifts for galaxies with
$1.62 < z < 1.65$ within a physical projected radius on the sky of
1~Mpc of the cluster center (10 of these galaxies have $1.62 < z <
1.63$; Papovich et al.\ 2010, Tanaka et al.\ 2010, I.~Momcheva et al.,
in prep, C.~N.~A.~Willmer et al., in prep).  These redshifts provide
an estimate of the velocity dispersion and total cluster mass assuming
the cluster is virialized, $\sigma_V = 360$~km s$^{-1}$ and $M_{200}
\approx 2 \times 10^{13}$~\msol, although there is evidence to suggest
the assumption of virialization is unlikely
\citep[see][]{papo10a,pierre11}, which is entirely consistent with
the expected assembly histories of a present-day massive cluster
observed at $z\sim 1.6$ \citep[\eg,][]{boylan09}.  While the reported
velocity dispersion was consistent with the weak ($4\sigma$)
\textit{XMM} X--ray detection \citep{papo10a},  recent
\textit{Chandra} data show that several point sources dominate the
X-ray emission with very faint extended emission, supporting the
interpretation that this cluster is in the act of collapsing
\citep{pierre11}.

In terms of semantics, throughout this paper we refer to \fulltarget\
as a ``cluster'' even though it is unlikely to fully satisfy the
classical definition of a virialized object.  The distinction
``proto-cluster'' or ``forming cluster'' is strictly more apt as it
seems likely that this structure is in the process of collapse and
assembly.    Regardless, because this object corresponds to a clear
high surface density of galaxies at $z=1.62$ \citep[20$\sigma$ as
defined by][]{papo10a}, we have the ability to compare and contrast
the morphological evolution of galaxies in a high density region
compared to that in the lower density field.

The outline for this paper is the following.    In \S~2 we describe
the properties of the imaging datasets and we describe our analysis.
In \S~3 we discuss the color--morphology relation in this cluster.  In
\S~4, we discuss the size--mass relation for quiescent galaxies
associated with the cluster and compare it to a similarly-selected
sample in the field.  In \S~5 we discuss the distributions of
ellipticities and surface--brightness profiles for the quiescent
galaxies in both the cluster and field.  In \S~6 we consider possible
evolutionary scenarios for the quiescent galaxy population, and we
discuss how environmental processes affect the galaxies' evolution.
In \S~7 we summarize our conclusions.  Throughout this paper we report
magnitudes measured relative to the AB system \citep{oke83}.  We
denote photometric magnitudes measured in the WFC3 F125W and F160W
passbands as $\wfcj$ and $\wfch$, respectively.  Throughout we assume
a cosmology with $\Omega_m=0.3$, $\Omega_{\Lambda}=0.7$, and $H_0=70$
km s$^{-1}$ Mpc$^{-1}$.    

\section{Data and Analysis}

The $z=1.62$ cluster \fulltarget\ is located in the  UKIRT IR Deep Sky
Survey \citep[UKIDSS;][]{lawr07}  Ultradeep survey (UDS). This cluster
received partial \hst/WFC3 imaging in the F110W and F160W bands as
part of the Cosmic Assembly Near--IR Deep Extragalactic Legacy Survey
(CANDELS) program (PIs: S.~Faber,
H.~Ferguson)\footnote{http://candels.ucolick.org/}.  The CANDELS
strategy, data acquisition, and data reduction are described fully in
\citet{grogin11} and \citet{koek11}. The CANDELS imaging achieves limiting
magnitudes of $\wfcj=\wfch=26.6$ mag ($10\sigma$ for apertures of
0\farcs4 diameter).    Owing to the CANDELS field placement, the
\hst/WFC3 imaging covers slightly more than 50\% of the galaxies associated
with the $z=1.62$ cluster (see below), including 6 galaxies with
spectroscopic redshifts $1.62 < z < 1.65$ within a physical projected
radius of 1~Mpc of the cluster center.  The CANDELS imaging does cover
most of the cluster core including its most massive, quiescent
galaxies. 

In addition to the \hst\ imaging, this field has deep $BRiz$ imaging
from the Subaru--XMM Deep Survey \citep[SXDF;][]{furu08}, $JK$ imaging
from UKIDSS \citep[\eg,][]{will09},  \spitzer\ IRAC data in four bands
probing $3.6$ to $8.0$~\micron, and MIPS data at
24~\micron.\footnote{http://irsa.ipac.caltech.edu/data/SPITZER/SpUDS}

\subsection{Merged Catalogs, Photometric Redshifts, and Sample Selection}

As in \citet{papo10a}, we used the $K$--band
selected, SXDF and UDS catalogs from \citet{will09} and merged these
with the \spitzer/IRAC data.   Following Papovich et al., we used the
muliwavelength photometry to derive photometric redshift
probability distribution functions, $P(z)$, for each source using EAZY
\citep{bram08}.  Here we considered a sample with $J \leq 24.5$~mag,
which is an approximate $3\sigma$ limit for the UKIDSS data.  As in
\cite{papo10a}, we define a likelihood that galaxies are associated with
the cluster redshift, 
\begin{equation}
\mathcal{P}_z \equiv \int\,\, P(z)\, dz, 
\end{equation}
integrated over the redshift range given by $z=z_\mathrm{cen}
\pm \delta z$ with $z_\mathrm{cen} = 1.625$ and $\delta z = 0.05
\times (1+ z_\mathrm{cen})$, approximately the 68\% confidence range
on the photometric redshifts for the red, quiescent galaxies.   

We consider all galaxies with $\mathcal{P}_z  > 0.3$ and projected
distances $R_\mathrm{proj} < 1.5$~Mpc to be associated with the
cluster.  Galaxies with well--established spectral features, such as
the 4000\AA/Balmer break, have sharp $P(z)$ and thus higher
$\mathcal{P}_z$, which includes red galaxies with lower implied
specific star--formation rates.    Galaxies that are actively
star--forming have weaker 4000\AA/Balmer breaks, have more broad
$P(z)$, and  have lower $\mathcal{P}_z$.  Therefore, choosing
$\mathcal{P}_z > 0.3$ ensures that we do not bias ourselves away from
the (bluer) star-forming objects \citep[see discussion in][]{papo10a}.
%

However, in the sections \S~4 and 5, we focus on the properties of a
sample of quiescent  galaxies in the cluster to those in the field.
For this sample of  quiescent galaxies we increase our selection
criterion to $\mathcal{P}_z  > 0.5$.    We do this because the
quiescent galaxies have tighter $P(z)$ functions, and will have higher
$\mathcal{P}_z$.  Our tests have shown that a $\mathcal{P}_z > 0.5$
criterion provides a cleaner sample as the samples  would otherwise
include galaxies with more than 50\% of their $P(z)$ outside the
desired redshift range.  From this subsample we define  quiescent
galaxies associated with the cluster as those with $R_\mathrm{proj} <
1.5$~Mpc, and those in the field as $R_\mathrm{proj} > 3.0$~Mpc.

Table~\ref{table} lists the properties of the objects in the CANDELS
cluster and field samples including the astrometric coordinates,
magnitudes, colors, and photometric redshift information.   The
table includes all galaxies at $z=1.6$ satisfying $\mathcal{P}_z >
0.3$ as defined above.

\subsection{Stellar Masses}

We fitted the 10--band galaxy photometry covering 0.4--8~\micron\ with
model spectral energy distributions to estimate the stellar masses for
the galaxies in the sample using the method of \citet{papo01}.  We
used models for a range of stellar population properties from the
\citet{bruz03} stellar population synthesis models allowing for a
range of extinction using the \citet{calz00} law.  We opt to use the
2003 version of the Bruzual \& Charlot models to facilitate the
comparison to other studies, including \citet{shen03}.  Our tests
showed that using the 2007 updated version of the Bruzual \& Charlot
models yields stellar masses systematically lower by a 0.2-0.3~dex,
but this does not affect our conclusions.  We assumed models with
solar metallicity and a Chabrier initial mass function (IMF; using a
Salpeter IMF would to first order increase systematically the stellar
masses by $\simeq$0.27 dex). Given that
most of the galaxies associated with the cluster are quite massive
\citep{tran10}, the solar--metallicity assumption is reasonable
\citep[see, \eg,][]{vandokkum04}.   Using different assumptions for
the stellar population metallicities will affect the derived stellar
masses by $\simeq$0.2~dex \citep{papo01,marc09}.  We generate a
multi-parameter probability distribution function for each galaxy from
this modeling.  We then compute the mean and 68\% confidence region on
the stellar mass for each galaxy by marginalizing over the other model
parameters \citep[see discussion in][]{papo06a}.  Our analysis of the
spectral energy distributions of  galaxies provides an estimate of the
instantaneous SFR, which we measure as the SFR averaged over the prior
100 Myr using the best-fit stellar-population model.
Table~\ref{table} lists the derived stellar masses and SFRs for each
object in the sample.

\subsection{Galaxy Morphologies and Sizes}

We used GALFIT \citep{peng02} to fit models to galaxies in the CANDELS
WFC3 F125W imaging, from which we determined effective radii,
$R_\mathrm{eff}$, and \sersic\ indices, $n$.  The models assume the
surface brightness of the galaxies is proportional to $\exp(-R /
R_\mathrm{eff})^{1/n}$ \citep{sersic68}, where $R$ is the angular
radius from the galaxy center and where the \sersic\ index is a
concentration parameter.  An exponential disk has $n = 1$, and a
\citet{devauc48} profile has $n = 4$.  GALFIT convolves the models by
the image PSF before fitting them to the data.     We generated model
PSFs for each dither position and orientation for the two WFC3 imaging
epochs using TinyTim v7.2 \citep{krist95}.  The PSF models were
dithered and combined in the same way as the CANDELS data.  

\ifsubmode
\begin{figure} 
\else
\begin{figure}  \epsscale{1.2} \fi
\plotone{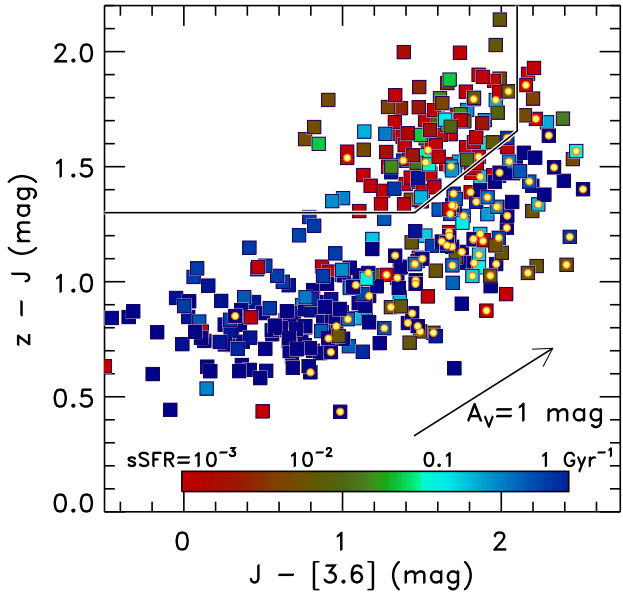}\caption{The observed $J - \mone$ versus $z -
J$ color--color diagram for galaxies at $z=1.6$ in the CANDELS UDS
field.   At $z=1.6$ these observed colors correspond approximately to
$V - J$ and $U - V$ rest--frame, which is very effective in separating
quiescent and star--forming galaxies \citep{will09}.  Quiescent
galaxies are expected to populate the upper left region of the plot
denoted by the polygon.  Star-forming galaxies form a sequence below
the quiescent region, where the arrow illustrates the expected change
in color for $A(V) = 1$~mag of dust extinction for a galaxy at
$z=1.6$.   The symbol colors scale with the specific SFR (sSFR, the
SFR per unit stellar mass) as defined by the inset color bar.  Small
yellow circles denote 24~\micron-detected sources with
$f_\nu(24\micron) > 50$~\ujy.}\label{fig:cc}  \epsscale{1} \ifsubmode
\end{figure}
\else
\end{figure}
\fi

\ifsubmode
\begin{figure} 
\else
\begin{figure*}  \epsscale{1.15} \fi
\plotone{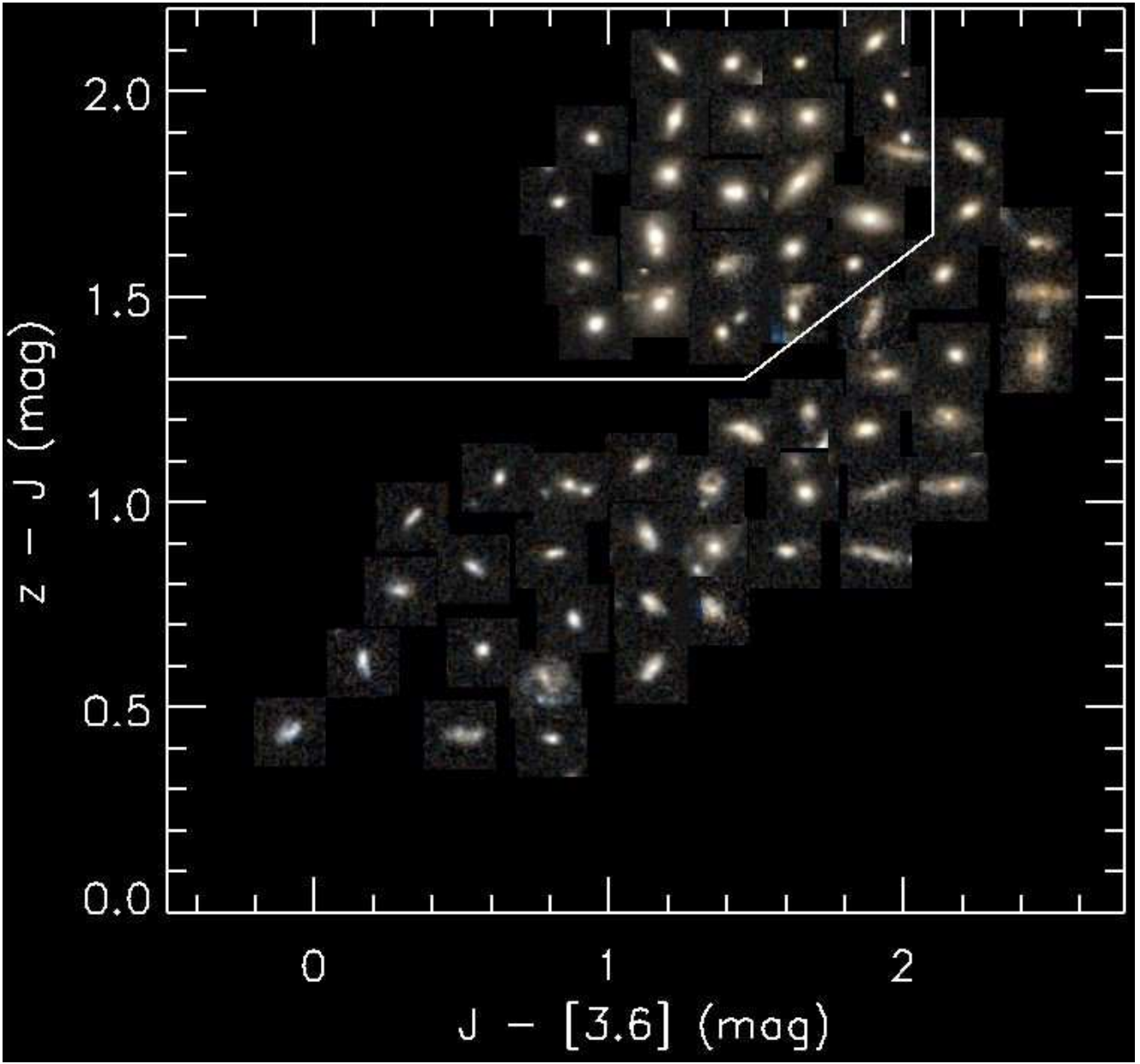}\caption{The observed $J - \mone$ versus $z -
J$ color--color diagram for galaxies associated with the $z=1.62$
cluster.  The plot includes all objects within 1.5 Mpc (projected) of
the cluster with $\mathcal{P}_z > 0.3$ (see text) and that have \hst\
coverage from CANDELS.  The color images show $6\arcsec \times
6\arcsec$ (approximately 50 kpc $\times$ 50 kpc at $z=1.6$) cutouts
from the CANDELS WFC3 F125W and F160W data.  The images are placed at
the approximate measured color each galaxy (slight adjustments to the
measured colors have been applied for clarity, but these shifts have
no affect on the conclusions).  There is a clear color--morphology
relation in the galaxies associated with this
cluster. }\label{fig:ccimg}  \epsscale{1} \ifsubmode
\end{figure}
\else
\end{figure*}
\fi

\ifsubmode
\begin{figure} 
\else
\begin{figure}   \epsscale{1.2} \fi
\plotone{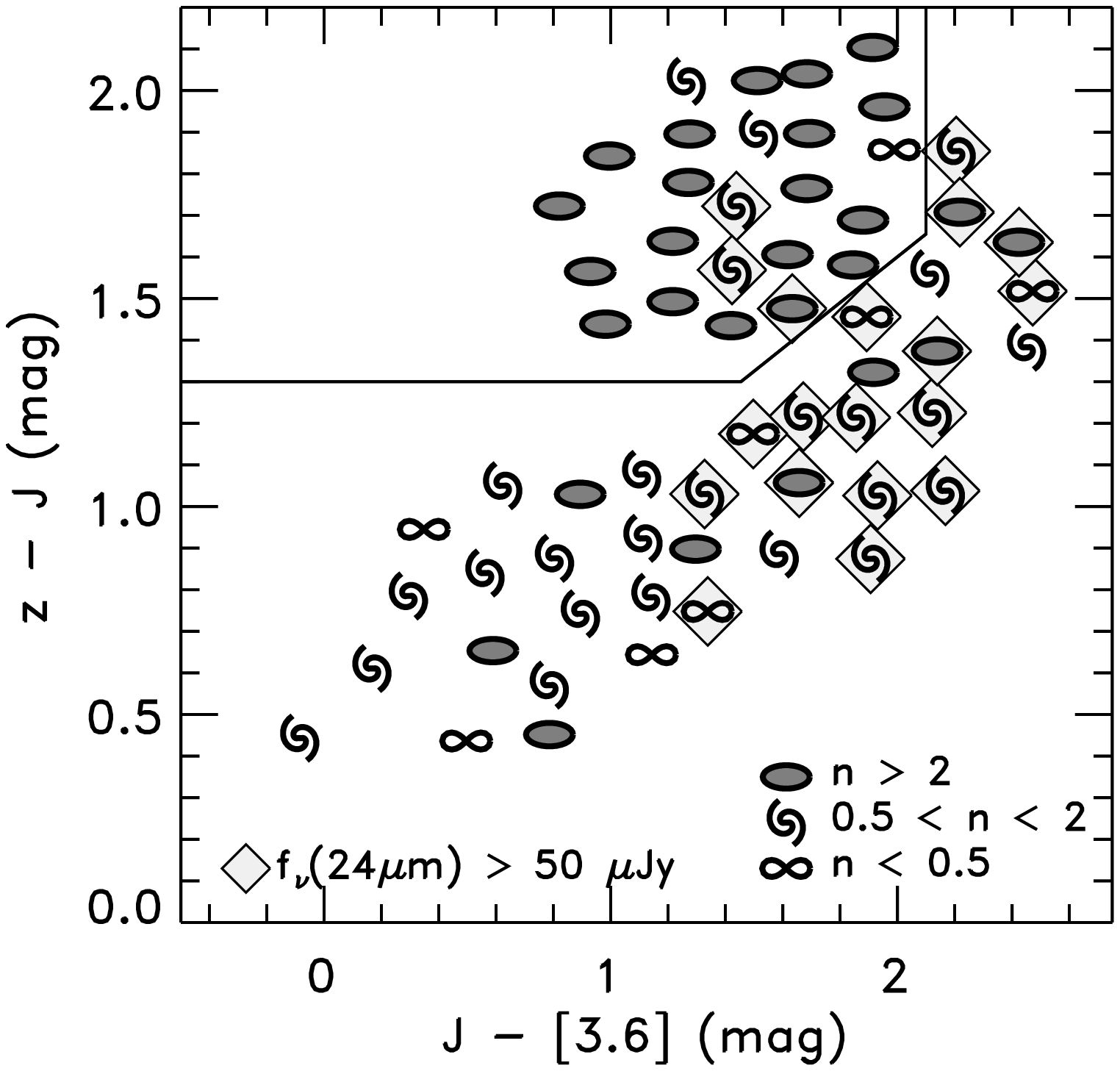}
\caption{$J - \mone$ versus $z - J$ color--color plot for galaxies
associated with the $z=1.6$ cluster.   The symbols denote the galaxy
\sersic\ indices, $n$, as labeled.  As in figure~\ref{fig:ccimg},   the
plot includes all objects within 1.5 Mpc (projected) of the cluster
with $\mathcal{P}_z > 0.3$ (see text) and that have \hst\ coverage in
CANDELS.  Also as in figure~\ref{fig:ccimg}, the data points are
placed at the approximate measured color each galaxy.  There is a
clear relation between the galaxies' morphological \sersic\ index and
their location in the color--color plane.  The galaxies in the
quiescent region of the plot have high \sersic\ indices, indicative of
galaxies with spheroid--dominated morphologies.  Galaxies with colors
of star-forming galaxies have lower \sersic\ indices, indicative of
disks and irregulars.  Sources denoted by gray diamonds have
24~\micron\ detections with $f_\nu(24\micron) > 50$~\uJy.
}\label{fig:ccSersic}  \ifsubmode
 \end{figure}
\else
\epsscale{1.0}
\end{figure}
\fi

We fitted each galaxy with GALFIT, keeping the position, background, orientation,
effective semimajor axis, \sersic\ index, and ellipticity as free
parameters.    We used the WFC3 F125W image for this analysis as this
bandpass corresponds approximately to the rest--frame $B$-band at
$z=1.6$, facilitating the comparison to other datasets.  Our
tests show that none of our conclusions would be strongly affected if
we instead used the WFC3 F160W image.   During the analysis we
required $n \leq 6$ because  higher values of $n$ usually do not
improve the fit, and the covariance between $n$ and the effective
radius leads to an overestimate of $R_\mathrm{eff}$ for larger $n$.
Only three objects in our samples had best fits with $n > 6$, and we
refit those objects forcing $n=4$.  In what follows, we analyze the
galaxies' ellipticities, defined as $\epsilon = 1 - q$, where $q=b/a$
is the ratio of the semiminor to semimajor axes calculated by GALFIT.
Table~\ref{table} lists the GALFIT measurements for all objects in the
samples.  

The effective sizes we report in this paper are the circularized
effective radii, $R_\mathrm{eff} = \sqrt{ab} = a_\mathrm{eff}
\sqrt{q}$, where $a_\mathrm{eff}$ is the effective semimajor axis
measured by GALFIT and other values are as above.  The circularized
effective radius is smaller than the effective semimajor axis, but it
is commonly used in the literature, and we use it here for comparison.
The effective semimajor axes can be computed using the information in
Table~\ref{table}.  Furthermore, we have checked that the circularized
effective radii from GALFIT are in good agreement with independent,
non-parametric measurements of the galaxy half--light radii computed
following the methods in \citet{lotz08}.  

We performed a series of simulations to estimate the errors in the
GALFIT parameters.  We inserted model galaxies of known effective
radius, \sersic\ index, and magnitude into the WFC3 F125W data, and we
recovered their parameters using GALFIT as described above.   As with
other studies \citep[\eg,][]{haus07}, we find that the errors in
effective radius and \sersic\ index are correlated strongly, with larger
uncertainties on effective radius for objects with larger \sersic\
indices.   Quantitatively, our simulations show that the measured
effective radii are accurate to better than 40\% for simulated compact
objects (measured $n=4$ and $r_\mathrm{eff} < 0.5$ arcsec) with
magnitudes typical of the faintest objects in our samples,
$m(\mathrm{F125W})$=23~mag.  Similarly, the measured \sersic\ indices
are accurate to better than 20\% for these objects.  The uncertainties
are substantially lower for brighter and less compact ($n < 4$)
objects \citep[similar to the findings of ][]{haus07}.   In practice,
the errors on the \sersic\ index have no substantive impact on our
conclusions. 

\section{The Color--Morphology Relation in a $z=1.62$ Cluster}

We select quiescent galaxies using a $J - \mone$ and $z-J$
color--color selection.  At $z=1.6$ these colors correspond
approximately to rest-frame $V - J$ versus $U - V$ at $z=1.6$, which
\citet{will09} showed effectively separates quiescent galaxies from
star-forming galaxies \citep[see also][]{wuyts09,patel11,quadri11}.
Figure~\ref{fig:cc} shows the $J - \mone$ versus $z-J$ diagram for all
galaxies in the CANDELS UDS field with with $\mathcal{P}_z > 0.3$.
The symbol colors denote the specific SFR (sSFR:  the SFR per unit
stellar mass).  Star--forming galaxies form a sequence below the
quiescent galaxies where the slope of the sequence roughly follows the
expected change in color associated with dust extinction.   

Quiescent galaxies lie in the upper left region of
figure~\ref{fig:cc}, as indicated by the polygon defined by
\begin{eqnarray}\label{eqn:colorselection}
(z - J)_\mathrm{AB}\bs&\ge&\bs 1.3\,\,\mathrm{mag}\nonumber \\
(J - \mone)_\mathrm{AB}\bs& \le&\bs 2.1\,\,\mathrm{mag} \\
(z - J)_\mathrm{AB}\bs&\ge&\bs 0.5 + 0.55 (J - \mone)_\mathrm{AB}. \nonumber
\end{eqnarray} 
We define samples of quiescent galaxies as satisfying all the color
criteria of equation~\ref{eqn:colorselection}.   Based on the analysis
of the galaxies' spectral energy distributions (\S~2.2), galaxies
selected using the color selection above in the CANDELS sample have
low specific SFRs.  We find
that 69/78 (88\%) of the quiescent galaxies have specific SFRs
$<10^{-2}$~Gyr (including all but one of the quiescent galaxies
associated with the $z=1.62$ cluster).  Therefore, the quiescent
galaxies selected by the color selection above have spectral energy
distributions  indicative of highly ``suppressed'' SFRs
\citep{kriek06}.  The MIPS 24~\micron\ data give an independent
measure of star formation or the presence of an AGN.  Few of the
quiescent galaxies are detected at 24~\micron:  only 3 out of 24
cluster galaxies and 6 out of 72 field galaxies have $f_\nu(24\micron)
> 50$~\ujy.  We do not reject these sources from the quiescent sample
because the source of the 24~\micron\ emission in these galaxies is
uncertain.  However, given the small number of 24~\micron\ sources,
our tests show that none of our conclusions would change if we did
remove these sources.  

The galaxies associated with the $z=1.62$ cluster exhibit
a clear color--morphology relation.   Figure~\ref{fig:ccimg} shows
the \hst\ WFC3 (F125 and F160W) color images for the cluster galaxies
with $\mathcal{P}_z > 0.3$ and projected
distances $R < 1.5$~Mpc from the cluster center. Spheroids dominate
the morphologies of the cluster galaxies
with colors of quiescent stellar populations defined by
equation~\ref{eqn:colorselection}.   In most
cases these  galaxies are highly symmetric with elliptical and
lenticular morphologies and a range of sizes.   Galaxies with colors
consistent with ongoing star formation have disk--like and irregular
morphologies.     In several cases galaxies in the star--forming
region  show evidence for multiple components, including apparent
bulge and disk morphologies.  This is especially visible in the
star--forming galaxies with redder $J - \mone$ colors, and these
galaxies appear to have large effective sizes compared to the bluer
star--forming galaxies. 

\ifsubmode
\begin{figure*}
\else
\begin{figure*}
\fi
\epsscale{1.12}
\plottwo{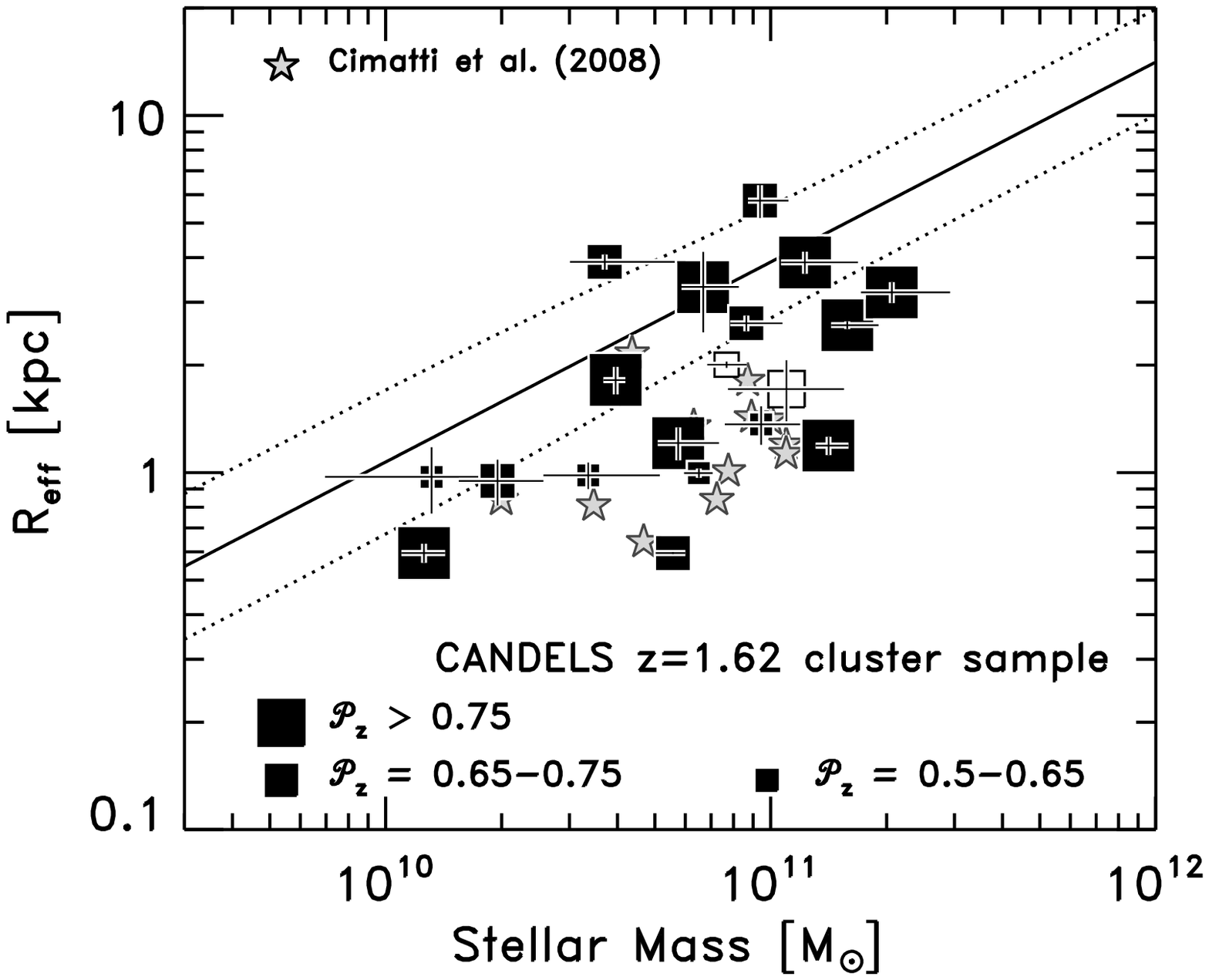}{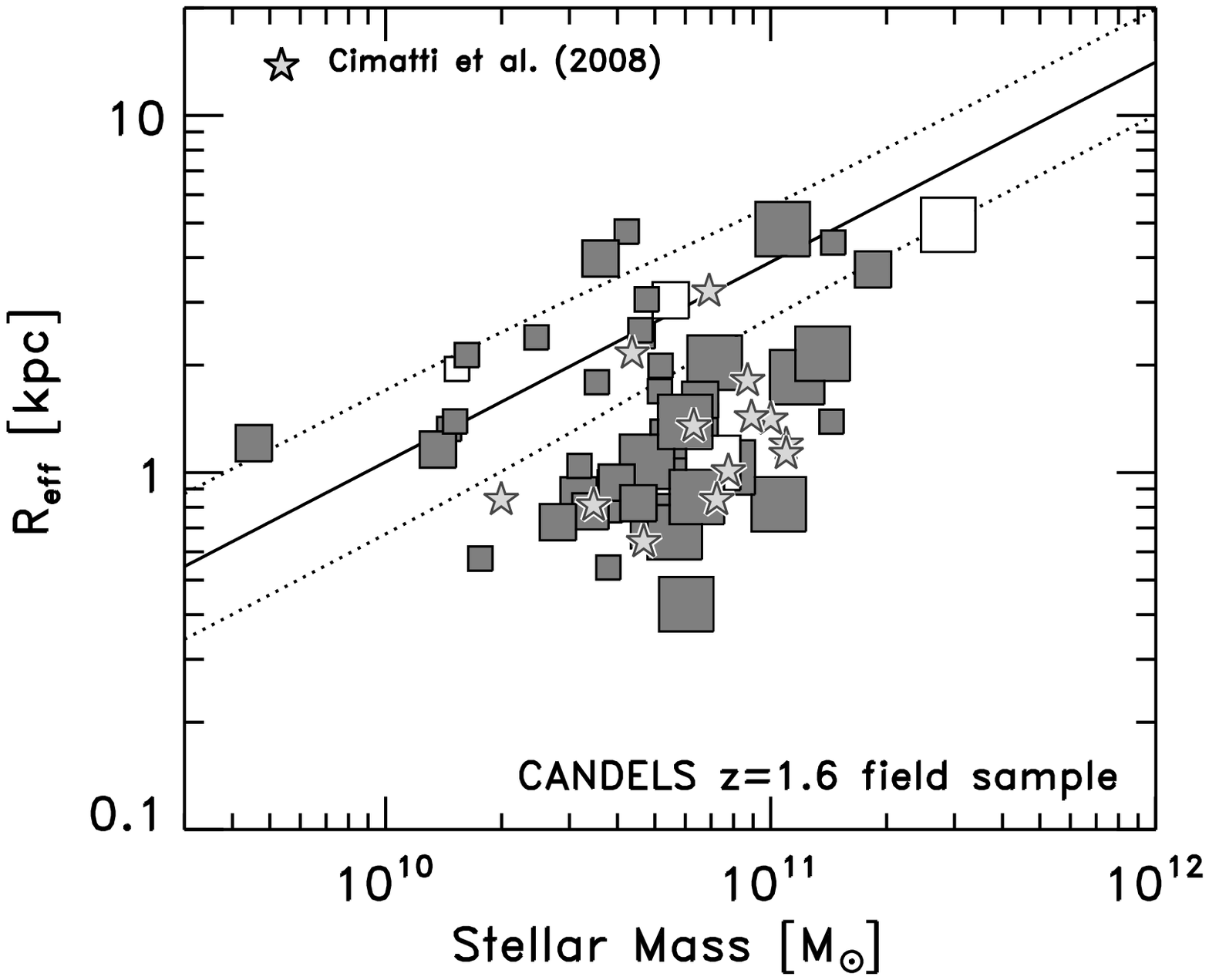}
\epsscale{1.0}
\caption{The left panel shows the relation between the circularized
effective radii and stellar mass for quiescent galaxies in the
$z=1.62$ cluster for galaxies with  projected distances
$R_\mathrm{proj} < 1.5$~Mpc from the cluster center and $\mathcal{P}_z
> 0.5$.    The right panel shows same relation for $z=1.6$ quiescent
galaxies in the field, selected in the same way as the cluster
galaxies but with $R_\mathrm{proj} > 3$ Mpc.  In both panels the size
of the data point (boxes) scales with, $\mathcal{P}_z$, as indicated
in the legend of the left panel.   The unfilled boxes denote objects
detected at 24~\micron\ with $f_\nu(24\micron) \geq 50$~\ujy.    The
solid and dotted lines show the $z=0.1$ size--mass relation for
early--type galaxies from the SDSS \citep{shen03}.  In each panel the
filled stars correspond to the $1 < z < 2$ early-type galaxy sample
from \citet{cima08}.   Quiescent galaxies in the field at $z\sim 1.6$ in the
CANDELS data have sizes similar to these other studies.   There is a
relative lack of compact quiescent galaxies in the cluster compared to
galaxies in the field.  } \label{fig:sizemass} \ifsubmode
\end{figure*}
\else
\end{figure*}
\fi
 
Figure~\ref{fig:ccSersic} shows the same $J - \mone$ versus $z - J$
color--color plot  as in figure~\ref{fig:ccimg} with the galaxies
denoted by symbols based on their \sersic\ indices as measured by
GALFIT.  Motivated by \citet{hogg04}, we classify galaxies with
high \sersic\ index, $n > 2$, low \sersic\ index, $0.5 < n < 2$, and very
low \sersic\ index $n < 0.5$.    Galaxies show a relation between their
\sersic\ indices and their location in the color--color plot of
figure~\ref{fig:ccSersic}.  Most of the galaxies with high \sersic\
indices fall  in the region of the plot occupied by quiescent
galaxies:  the quiescent galaxies have surface-brightness profiles
dominated by spheroids.    Galaxies with low and very low \sersic\
indices fall primarily in the region of the plot occupied by
star--forming galaxies:  they have surface-brightness profiles
dominated by disks.  Quantifying these statements, we find that of the
24 cluster galaxies with $z - J$ and $J - \mone$ colors of quiescent
galaxies, 19 (79\%) have $n > 2$, suggesting a high early--type galaxy
fraction among the passive galaxies in the cluster.  Of the galaxies
in the star--forming region of figure~\ref{fig:ccSersic}, 28 of 38
galaxies (74\%) have $n < 2$, implying they are dominated by objects
with disk--like or irregular morphologies.     Furthermore, based on
the simulations in \S~2.3 the errors on the \sersic\ index have no
substantive impact on our conclusions.  Therefore, the
color--morphology relation exists in this $z=1.62$ cluster, with high
\sersic\-index (spheroid-dominated) galaxies populating the quiescent
region of the color--color plot, and with low \sersic\-index galaxies
populating the star--forming region.  This extends a similar result
observed for field galaxies \citep{wuyts11} and \citep{bell11} to higher density
regions associated with the cluster at these redshifts. 

\section{The Size--Mass Relation for Quiescent  Galaxies at $\mathbf{z=1.6}$}\label{section:sizemass}

\subsection{Comparison between Cluster and Field Quiescent Galaxies}\label{section:sizemass1}

Figure~\ref{fig:sizemass} shows the (circularized) effective radii of
the the quiescent galaxy samples from CANDELS in both the $z=1.62$ cluster and the
$z=1.6$ field as a function of their stellar mass.  
%
%
At fixed stellar mass, the quiescent field galaxies at $z=1.6$ in the CANDELS
data generally have effective radii smaller by about a factor of 3
compared to the distribution of low redshift early-type galaxies from
SDSS \citep{shen03,guo09}, consistent with previous
results \citep[see,][and references therein]{cima08,cass10}.  

In contrast, the quiescent galaxies associated with the $z=1.62$
cluster show a relative lack of compact galaxies compared to the
quiescent field galaxies at $z=1.6$ at fixed mass.  Quantitatively,
the quiescent cluster galaxies with masses $>3\times 10^{10}$~\msol\
and $\mathcal{P}_z > 0.5$ have an interquartile (25--75\%-tile) range
of $R_\mathrm{eff} = 1.2 - 3.3$~kpc with a median of 2.0~kpc, whereas
the field galaxies have an interquartile range, $R_\mathrm{eff} =
0.9-2.4$~kpc with a median of 1.3~kpc.  The size of a typical massive,
quiescent galaxy in the $z=1.62$ cluster is larger compared to field
galaxies.  This trend is consistent qualitatively with recent findings
by \citet{cooper11} and \citet{zirm11}.     We note, however, that the
difference between the sizes of the cluster and field quiescent
galaxies declines at higher masses, as many of the galaxies with $M
\gsim 10^{11}$~\msol\ in both the field and cluster samples  have
larger effective radii \citep[$\gsim 2$~kpc; similar to the findings
of ][]{rettura10}.   We note that recent work from \citet{raic11}
concludes an opposite trend such that early-type galaxies in higher
density regions are smaller.  However, the significance of this result
is likely a consequence of sample selection and analysis method, as
discussed in \citet{cooper11}.  

The relative lack of compact quiescent galaxies in the $z=1.62$
cluster is unlikely a result of selection effects. There are inherent
biases and systematics in the measurement of both the effective sizes
and stellar masses \citep[see, e.g.,][]{papo01,papo06a,haus07}.
However, these mostly affect comparisons between samples of galaxies
taken from different datasets and at different redshifts.  In the case
here, both the cluster and field galaxy samples are selected at the
same redshift and using the identical CANDELS dataset.  Therefore, the
same systematics and biases affect both  samples equally.  As a
result, the  \textit{relative} comparison between the galaxies in the
cluster and field is robust.

\ifsubmode
\begin{figure}
\epsscale{1.0}
\else
\begin{figure}[b]
\epsscale{1.15}
\fi
\plotone{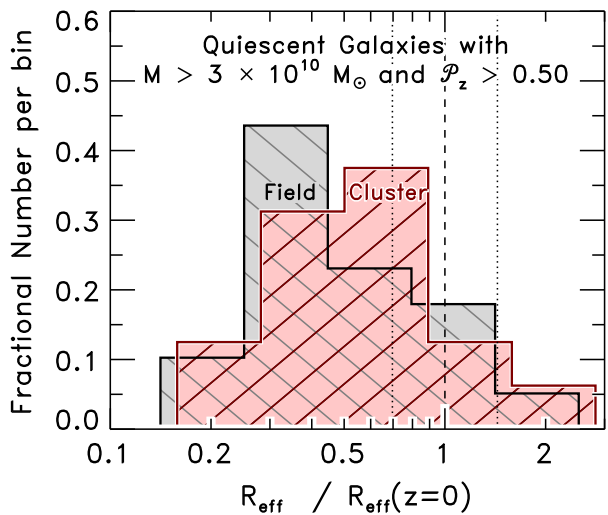}
\caption{Comparison of the distribution of circularized effective
radii of quiescent galaxies in the $z=1.62$ cluster and the $z=1.6$
field in the CANDELS UDS data.  The size distribution is measured
relative to local early--type galaxies of \citet{shen03}.   The
vertical dotted lines show the scatter about the mean relation (dashed
line) from \citet{shen03}.   The histograms show the distribution for
both the $z=1.62$ and cluster and $z=1.6$ field quiescent galaxies
with solar masses $>3\times 10^{10}$~\msol\ and $\mathcal{P}_z > 0.5$.
The average relative size of a quiescent galaxy is larger compared to
quiescent field galaxies.    This is primarily due to the lower number
of compact quiescent galaxies in the cluster.
}\label{fig:relsizehist} \epsscale{1.0}
\end{figure}

Figure~\ref{fig:relsizehist} shows the distribution of effective radii
for the $z=1.62$ cluster and $z=1.6$ field quiescent galaxies relative
to the low-redshift relation for early-type galaxies from
\citet{shen03}.   The CANDELS UDS samples include quiescent galaxies
with stellar masses $>3 \times 10^{10}$~\msol and $\mathcal{P}_z >
0.5$.   The main difference in the samples is that the cluster
galaxies at $z=1.62$ have a relative lack of quiescent galaxies with
low effective sizes compared to the field sample, as discussed above.
Formally, a Mann-Whitney-Wilcoxon rank-sum test \citep{mann47} gives a
12\% likelihood ($\simeq 1.2\sigma$)  that both the CANDELS $z=1.62$
cluster and $z=1.6$ field samples are drawn from the same parent
sample.   The significance increases to $\simeq 2\sigma$ if we
consider a higher fidelity sample of cluster and field galaxies with
integrated photometric redshift probability distribution
$\mathcal{P}_z > 0.65$.  

There is strong evidence that the size distribution evolves from $z=
1.6$ to $z\sim 0$, as inferred from other studies.   We test this
by computing a likelihood that the effective sizes of
quiescent field galaxies at $z=1.6$ and quiescent cluster galaxies at
$z= 1.62$ have the same mean sizes as the local sample of early-type galaxies
from \citet{shen03}.   A Student's $t$-statistic gives likelihoods of
$2 \times 10^{-7}$ and $6\times 10^{-3}$, for the $z=1.6$ field and
$z=1.62$ cluster samples respectively,  where the higher
likelihood for the cluster  is a result of the fact that the mean size
of the cluster galaxies is larger than that of the field galaxies.
Regardless, based on these tests both the quiescent galaxies in the
cluster and field at $z=1.6$ show strong size evolution compared to
the field. 

\subsection{The Evolution of the Size-Mass Relation in Clusters}

Figure~\ref{fig:relsizes} shows the evolution of the relative mean
sizes for quiescent galaxies for CANDELS in the $z=1.62$ cluster and
in the $z\sim 1.6$ field from the compared to other samples in the
literature.  In all cases we measure the size relative to the
low-redshift relation from SDSS \citet{shen03}.  The figure includes the mean relative sizes of early--type
galaxies in other high-redshift cluster samples, including MS 1054-03
at $z=0.83$ and RX J0152.7-1357 at $z=0.83$ from \citet{blak06} and
\citet{holden09}, and RDCS 1252.9-2927 at $z=1.24$ from
\citet{rettura10}.   In addition, the figure shows the mean relative
sizes for the field sample of early-type galaxies of \citet{cima08} at
$z\sim 1.4-2.0$.    The shaded curve shows the best-fit relation to
the evolution of quiescent galaxies from \citet{vanderwel08}. 

We parametrize the size evolution in figure~\ref{fig:relsizes} for
the cluster samples as $R_\mathrm{eff} \propto (1+z)^\alpha$.  Fitting
the data points for the clusters of \citep{blak06} and
\citet{rettura10}, and the $z=1.62$ cluster from the CANDELS data,  we
find $\alpha = -0.6 \pm 0.1$.   This is highly consistent with the
30\% increase in the sizes of brightest cluster galaxies at $z=1$ to
0.25 measured by \citet{stott11}.   In comparison, \citet{vanderwel08}
derived a steeper exponent, $\alpha = -1.0 \pm 0.1$ considering
samples of field and cluster early-type galaxies, and this rapid
evolution seems required to match the mean sizes of the very compact,
passive galaxies at $1 < z < 3$ \citep[see also,][]{damj11}.     The
data for the quiescent galaxies in the $z=1.62$ cluster suggest that
quiescent galaxies in the high density region of clusters experience
slower size evolution from $z= 1.6$ to $z\sim 0$ compared to the
field. 


\section{Evidence for Stellar Disks in Quiescent Cluster
  Galaxies}\label{section:disks}

\subsection{Ellipticity Distributions}\label{section:ellipticity}

As discussed in \S~3, the quiescent galaxies in the $z=1.62$ cluster
have concentrated, spheroid-dominated morphologies (\sersic\ indices $n >
2$).  In addition, many of the quiescent galaxies have low axial
ratios, corresponding to high ellipticities ($\epsilon = 1 - b/a$
where $b/a$ is the ratio of the semi-minor to semi-major axes from
GALFIT, see \S~2.3).  Indeed, many of the quiescent galaxies
in figure~\ref{fig:ccimg} show elongated morphologies with
significant ellipticity.   

\citet{vanderwel11} recently cited the high ellipticities of a
majority of field quiescent galaxies at $z\sim 2$ as evidence that
these galaxies have prominent disk components, consistent with other
studies \citep{mcgrath08,weinzirl11}.  This is similar to the
observation for both the cluster and field galaxies at $z=1.6$ here.

Figure~\ref{fig:b2a_mass} shows the measured ellipticities for the
quiescent galaxies in the $z=1.6$ field and the $z =1.62$ cluster as a
function of stellar mass.   Both the quiescent galaxies in the
$z=1.62$ cluster and $z=1.6$ field in the CANDELS data have relatively
high measured ellipticities.  The median ellipticity for both samples
$\epsilon_\mathrm{med} = 0.4$.  As illustrated in
figure~\ref{fig:b2a_mass}, there is no strong evidence that the
ellipticity distributions differ between the $z=1.6$ field and
$z=1.62$ cluster samples.    A Mann-Whitney-Wilcoxon rank-sum test
finds no statistically significant difference between the ellipticity
distributions for the two samples:  we are unable to reject the
hypothesis that they are drawn from the same parent sample.  

\ifsubmode 
\begin{figure}  
\epsscale{1.} 
\else 
\begin{figure}[t]
\epsscale{1.210} 
\fi
\plotone{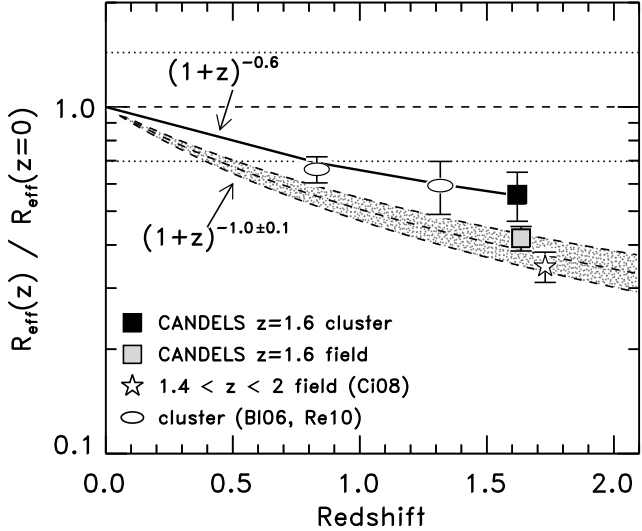}
\caption{  Comparison of the evolution of the effective radii for
early-type and quiescent galaxies.  The large squares show the
relative sizes of CANDELS quiescent galaxies in the $z=1.62$  cluster
(solid black datum) and the $z=1.6$ field  (solid gray datum).  The
ellipses show the mean relative sizes of early-type galaxies from two
clusters from \citet[Bl06]{blak06} using the stellar mass relation from
\citet{holden09} at $z=0.83$, and from one cluster at $z=1.24$
\citet[Re10]{rettura10}.  The stars show passive galaxies in the field
at $1.2 < z < 2.0$ \citep[Ci08]{cima08}.    All data points show the
galaxies sizes relative to the local size--mass relation of
\citet{shen03}.  The horizontal lines show this local relation its
scatter.   The shaded curve shows the size evolution in early--type
galaxies measured by \citet{vanderwel08} for a mix of field and
cluster galaxies.  The thick, solid line shows the size evolution
measured here for the cluster galaxies only.  The fit suggests milder
size evolution from $0 \lsim z < 1.6$ for cluster galaxies compared to
\citet{vanderwel08}. }\label{fig:relsizes}
\epsscale{1.0}
\end{figure}
 
The measured ellipticities of the quiescent galaxies at $z=1.6$ are
comparable to the ellipticities measured for lenticulars and
early-type spirals in clusters at $0 < z < 1$, which have
$\epsilon_\mathrm{med}^\mathrm{(S0)} = 0.4-0.5$\citep[\eg,][]{vulc11}.
In contrast, massive cluster ellipticals at low redshift have lower
ellipticities, $\epsilon \simeq 0.2-0.3$ \citep{holden09,holden11}
with no indications of evolution \citep[\eg,][]{vulc11}.  

However, unlike galaxy samples at lower redshifts
\citep[\eg,][]{holden09,vanderwel09b}, we find no evidence for a trend
between the ellipticity and stellar mass in the $z=1.6$ field and
cluster samples.    As illustrated in figure~\ref{fig:b2a_mass} the
ellipticities of the higher mass ($>10^{11}$\msol) galaxies in both
the $z=1.6$ field and $z=1.62$ cluster remains high, with a median
$\epsilon_\mathrm{med} = 0.4$ and with an interquartile range spanning
$\epsilon = 0.3-0.7$.  In contrast, \citet{vanderwel09b} find  a
median ellipticity of $\epsilon \simeq 0.2-0.3$ for non-star-forming
galaxies at $0.04 < z < 0.08$ with $M > 10^{11}$~\msol, with no
apparent evolution to $z\sim 0.6-0.8$  \citep{holden11}.  Therefore,
the massive ($>10^{11}$~\msol) quiescent galaxies in both the $z=1.62$
cluster and $z=1.6$ field have  higher ellipticities than lower
redshift ($z \lsim 1$) counterparts. 

\subsection{Surface Brightness Profiles of Quiescent Cluster Galaxies}

These ellipticities may indicate flattened disk--like structures
viewed in projection.  Roughly 50\% of  the cluster sample, and 30\%
of the field sample have $\epsilon  > 0.5$ (see
figure~\ref{fig:b2a_mass}).  Assuming inclination angles are
distributed randomly, this implies that a large portion of the
massive quiescent galaxies have disk components \citep{lambas92}.   We
investigate the presence of disk components by studying the surface
brightness profiles of the four most massive quiescent galaxies
associated with the $z=1.62$ cluster, IDs 39716, 40170, 40640, and
42952 (see Table~1).     These four galaxies all have stellar masses
$>1\times 10^{11}$~\msol\ (see figure~\ref{fig:sizemass}) with low
levels of star formation.\footnote{ We find from the analysis of their
spectral energy distributions (\S~2.2 and 3) limits on the SFRs of
$\Psi < 10$~\msol\ yr$^{-1}$, with the exception of 42952 which is
consistent with $\Psi \lsim 40$~\msol\ yr$^{-1}$.  These SFRs are
consistent with the limits from their (lack of) detected \spitzer\
24~\micron\ emission, $f_\nu(24\micron) < 40$~\uJy, implying
$\Psi(24\micron) < 5$~\msol\ yr$^{-1}$.  Including both the
constraints from the 24\micron\ data and analysis of spectral energy
distribution, the specific SFRs for these galaxies are very low, $<
5\times 10^{-2}$~Gyr$^{-1}$.}  Three of these four most massive
galaxies (39716, 40640, 42952)  have $\epsilon > 0.4$. 
%

Figure~\ref{fig:sbfit} shows the one--dimensional surface brightness
profiles for these galaxies.   We fit each galaxy using three models.
These include a best--fit GALFIT model using a single component with
the \sersic\ index, $n$, as a free parameter.  We also considered a
model with a single component  with the \sersic\ index fixed at $n=4$.
Lastly, we considered a model with two components,  where the \sersic\
index is fixed at $n=4$ for one component and at $n=1$ for the other
component.    In addition, objects 39716 and 40170 show indications of
faint companions with angular separations of less than one arcsecond.
For the analysis here, we masked the light from these objects to
prevent them from affecting these surface brightness fits.   However,
we find that masking these faint objects changes the derived effective
sizes and ellipticities by $<$15\%.

\ifsubmode
\begin{figure}
\epsscale{1.0}
\else
\begin{figure}[t]
\epsscale{1.2}
\fi
\plotone{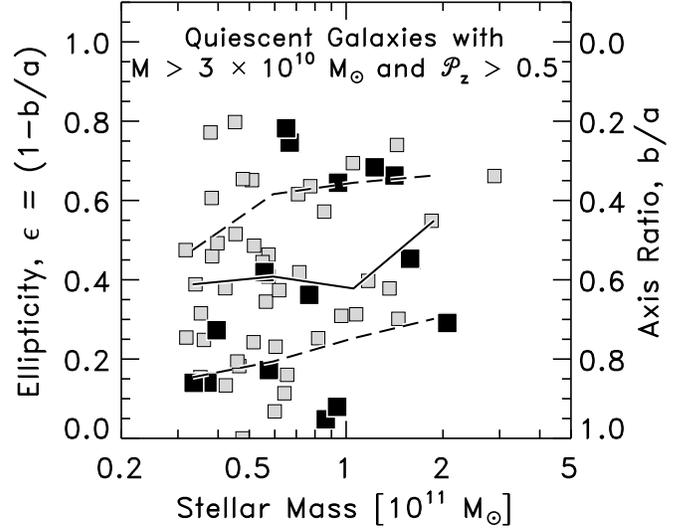}
\caption{ The  distribution of galaxy ellipticity, $\epsilon =
(1-b/a)$, as a function of stellar mass for the quiescent galaxies in
the $z=1.62$ cluster (filled boxes) and in the $z=1.6$ field (lightly shaded
boxes) in the CANDELS UDS data. There is no statistically significant
difference between the distributions for the cluster and field
samples.  The lines show the median and interquartile (25-75\%-tile)
values in bins of 0.25 dex in mass for the combined field and cluster
sample.  The median ellipticity, $\epsilon \simeq 0.4$, at stellar
masses $<10^{11}$~\msol\ is similar to values found for
non-starforming galaxies in SDSS at $0.04 < z < 0.08$
\citep{vanderwel09b}.  However, there is no strong trend between the
ellipticity and stellar mass, which contrasts with observations of lower redshift galaxies \citep[\eg,][]{holden09,vanderwel09b}.   }\label{fig:b2a_mass}
\end{figure}

\ifsubmode
\begin{figure} 
\plottwo{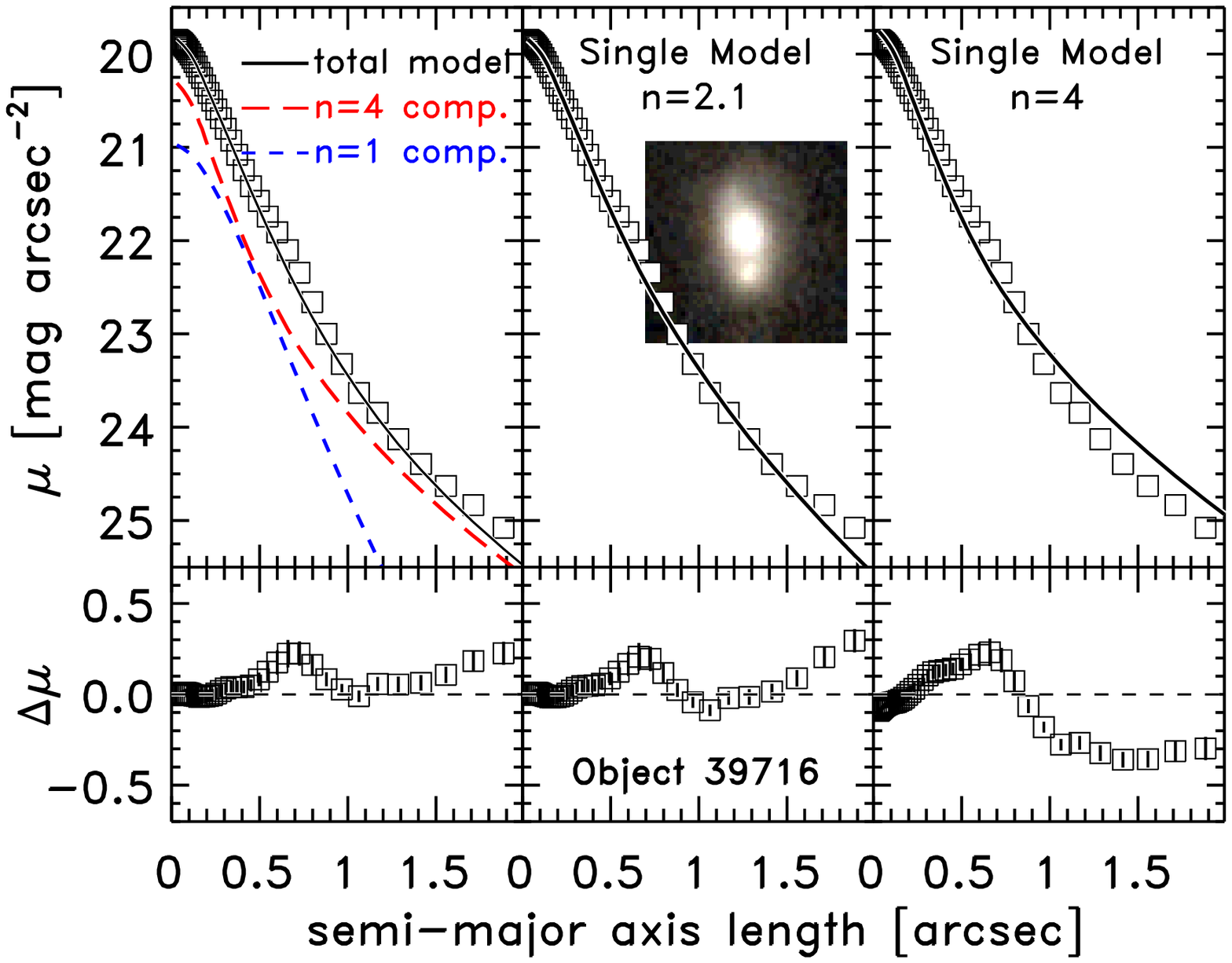}{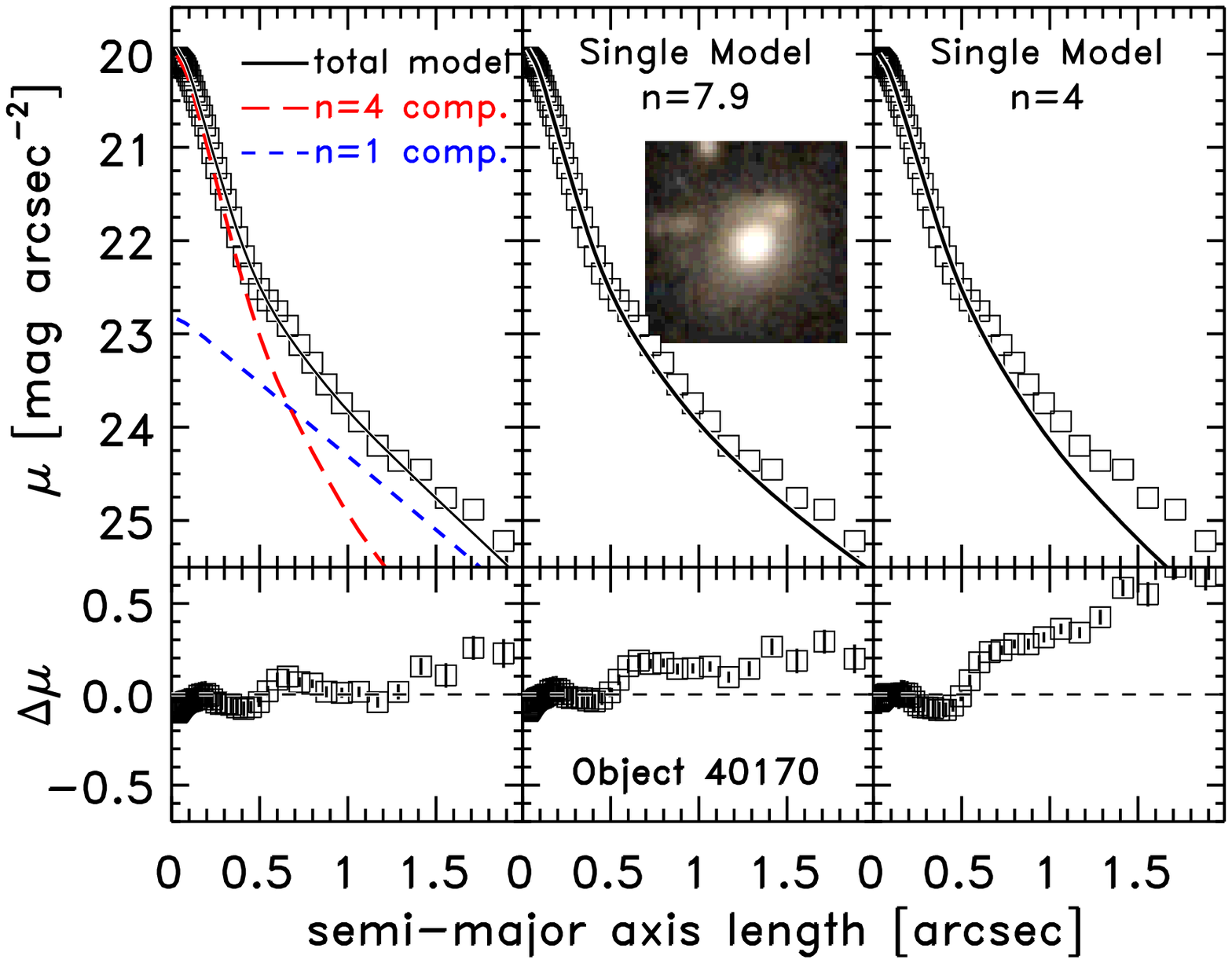}
\plottwo{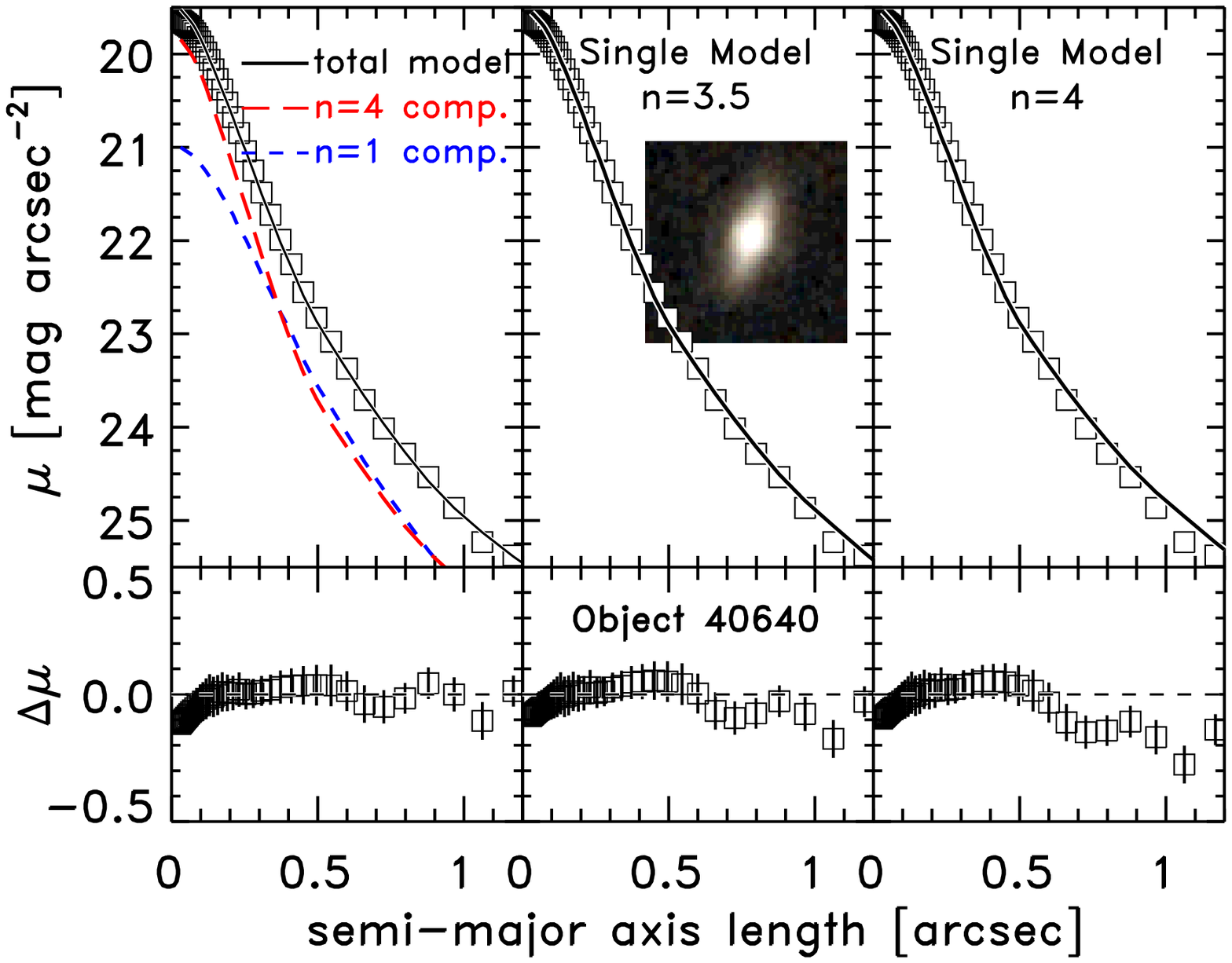}{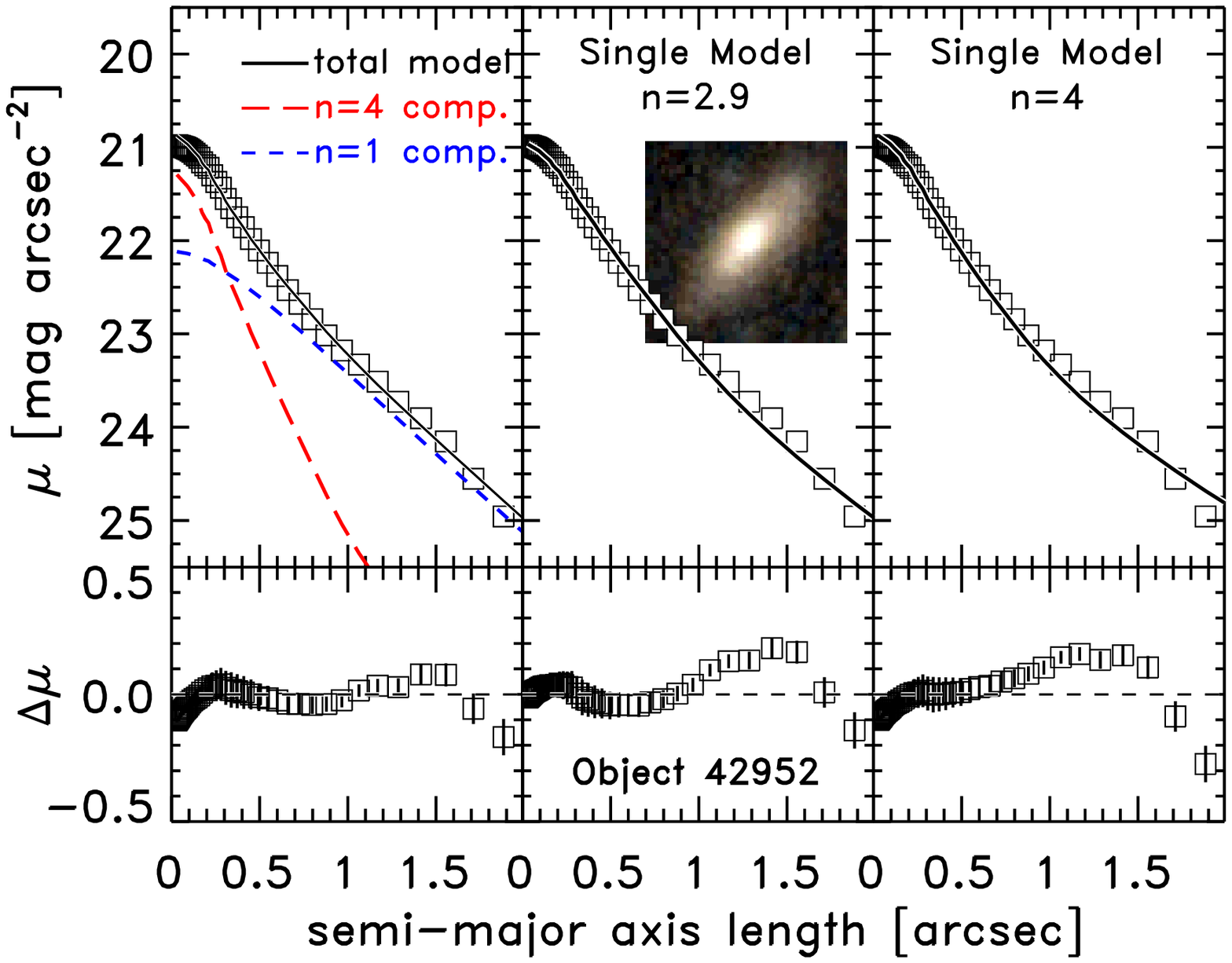}
\else
\begin{figure*}
\epsscale{1.05}
\plottwo{fig8a.ps}{fig8b.ps}
\plottwo{fig8c.ps}{fig8d.ps}
\fi
\epsscale{1.}
\caption{ Surface--brightness profiles for the four most massive,
quiescent galaxies ($M > 1\times 10^{11}$~\msol) associated with the
cluster at $z=1.62$, with ID numbers as labeled.  The box points in the top
panels show the measured surface--brightness profile.   The inset
images show a $6\arcsec \times 6\arcsec$ cutout of the galaxy using
the WFC3 F125 and F160 data.   
%
%
The curves in the panels show different model fits to the F125W data.
For each galaxy, the left panel shows a model with two--components,
where one model has a fixed \sersic\ index, $n=4$, and the other has a
fixed \sersic\ index, $n=1$.   The middle panel has a single model where
the \sersic\ index is a free parameter.  The right panel has a single
model with fixed \sersic\ index $n=4$.  The bottom panels for each
galaxy show the difference between the measured surface--brightness
and each model.  Error bars on the surface-brightness measurements are
shown in the bottom panels only for clarity. }\label{fig:sbfit}
\ifsubmode
\end{figure}
\else
\end{figure*}
\fi

In all cases the single component fits require \sersic\ indices $n > 2$
for these objects.  They are spheroid dominated.  Generally, the two
component models have lower residuals  between the model and the data,
particularly at larger radii (see the bottom panels for each galaxy in
figure~\ref{fig:sbfit}).  The disk exponential scale length for the
$n=1$ components range from 2--5~kpc, consistent with the disk scale
lengths for low redshift galaxies of comparable stellar mass in SDSS
\citep{fathi10b}.  In all cases the $n=1$ components have
ellipticities that are in within 20\% of the ellipticity from the fits
for each object with single-components.  The implication is that the
light profiles of these massive $z=1.62$ cluster galaxies are
inconsistent with a model of constant \sersic\ index, and they instead
favor a model with a radially dependent \sersic\ index to describe their
structure, with high \sersic\ indices at small radii changing to lower \sersic\
indices at larger radii. 

We use the two-component models to estimate crudely the
``bulge--to--total'' ($B/T$) ratio for each galaxy, defined as the
ratio of the flux in the $n=4$ component to the total flux.    Objects
39716, 40170, and 40640 are bulge--dominated, with $B/T = 0.5-0.7$,
consistent with those of lenticular galaxies and ellipticals \citep{simien86,graham08}.  Object
42592 has a lower ratio, $B/T = 0.3$.  The  disk component dominates
the light in this galaxy, similar to middle-to-early--type spirals
\citep{simien86,graham08}. Assuming the stellar mass traces the \wfcj\ light,
the derived $B/T$ imply that 30-50\% of the stellar mass lies in the
disk component for most of these galaxies, although the disk in 42592
may contain as much of 70\% of the stellar mass.  Interestingly,
unlike the other three objects (39716, 40170, 40640) with all reside
within 0.5~Mpc of the luster, object 42952 sits at a projected
distance 1.2~Mpc, and there is some evidence quiescent galaxies at
this distance exhibit more disk-dominated morphology (R.~Bassett et
al.\ 2012, in prep).

We conclude that these massive quiescent galaxies associated with the
cluster show evidence for prominent stellar disk components in their
surface brightness profiles.   The lack of significant star formation
in these galaxies suggests that  these disks are primarily stellar
systems.   However, it may also be the case that the spheroids form
through the migration of stars formed from violent instabilities in
the disk, which stabilizes the gas in the disk against further
instabilities that would otherwise form stars
\citep[\eg,][]{dekel09b,martig09}.      Currently,  the only
\hst--quality data (FWHM $\simeq 0.1-0.2\arcsec$) for these galaxies
is the CANDELS F125W and F160W imaging used here, and we are unable to
test for surface-brightness gradients indicative of variations in the
stellar populations of these possible bulge and disk
components. However, we see no measurable color gradients in WFC3
$\wfcj - \wfch$ images.  While this is consistent with
\citet{mcgrath08} who find negligible ACS $I_{814}$ - NICMOS $\nich$
color gradients of early--type galaxies at $z\sim 1.5$, \citet{guo11}
and \citet{szom11} find evidence that some quiescent galaxies at
$z\sim 2$ have negative color gradients with bluer cores and redder
outer regions.  To test for color-gradients in the galaxies here will
require data into the rest-frame near-UV using, e.g., \hst/ACS
observations. 

If this result generalizes to the full sample of quiescent cluster galaxies, the
ellipticity distribution in figure~\ref{fig:b2a_mass} provides evidence
that a large fraction of these galaxies host stellar disks.  The lack
of color gradients in WFC3 $\wfcj - \wfch$ images suggests the stellar
populations in the disk and spheroid components are fairly homogeneous
\citep[see, \eg,][]{papo05,mcgrath08}.    However, given that massive cluster
galaxies at low redshift are dominated by spheroids with $n \gsim 4$
and show no evidence for disks \citep{post05,holden09}, these disk
structures must be destroyed at some later time.  We discuss
the implications of this evidence below.

\section{Discussion}

\subsection{Accelerated Evolution in High Density Regions}

The quiescent galaxies in the $z=1.62$ cluster and the $z=1.6$ field
share many common properties. Their morphologies show dominant
spheroidal components.  However, both the field and cluster samples
have broad ellipticity distributions (figure~\ref{fig:b2a_mass} and
\S~\ref{section:disks}), suggesting the presence of disks.  Based on
the modeling of the galaxies spectral energy distributions and (lack
of) IR emission, these quiescent galaxies have low levels of star
formation (see \S~2.2 and 3), implying they either have low cold--gas
fractions, or that the cold gas in the galaxies is stable against
instabilities, perhaps as a result of the dominant spheroids
\citep{dekel09b,martig09}, or the dominance of a stellar component in
the disk \citep{cacc11}. 

The main difference between the cluster and field quiescent galaxies
at $z=1.6$ is the relative lack of compact, massive quiescent galaxies
in this cluster compared to those in the field
(figure~\ref{fig:sizemass}), and this is
significant at the $\gsim 90\%$ level
(\S~\ref{section:sizemass1}).  If correct, then this
result implies that the quiescent galaxies in the $z=1.62$ cluster
have experienced accelerated size growth relative to the quiescent
galaxies in the field at $z=1.6$. 

One possibility is that prior to $z=1.6$ the quiescent galaxies in the
cluster experience accelerated spheroid growth associated with the
migration of stellar clumps originally formed in the galaxy disks.
Theoretical considerations and cosmological hydrodynamical simulations
predict that galaxies at $z \gsim 2$ form stars from gravitational
instabilities in the disks fed by cold gas accreted in streams and
minor mergers along filaments \citep[]{dekel09a,cever10}.  Over $\sim
0.5$~Gyr the instabilities and clumps in the disk migrate inwards and
merge and form a passive spheroid \citep{dekel09b,bour11}.  The
spheroid stabilizes the disk (leaving it intact) against
instabilities, which suppresses star formation.  One feature of this
model is that spheroid-dominated galaxies at $z\sim 1.5-2$ should show
disk components with a scale length comparable to that of the spheroid
\citep{martig09}.  This is qualitatively consistent with our CANDELS observations of
quiescent galaxies in both the $z=1.62$ cluster and the $z=1.6$ field.
Nevertheless, because the sizes of the quiescent galaxies in the
$z=1.62$ cluster are larger on average compared to those in the
$z=1.6$ field, this would imply that processes of star-formation in
the disk and the migration of stars into a central spheroid happen
earlier and/or occur at an accelerated rate for the cluster galaxies,
possibly as a result of enhanced gas accretion associated with the
higher density region.   

Parenthetically, these simulations predict that a small fraction of
the gas that flows into the spheroids fuels the growth of supermassive
black holes (SMBH).  \citet{bour11} predict that at $z\sim 2$ a galaxy
with $10^{11}$~\msol\ will have a SMBH with an accretion rate that
corresponds to an X-ray luminosity of $10^{42-43}$~erg s$^{-1}$.  One
of the most massive quiescent galaxies in the $z=1.62$ cluster, ID
39716 (see Table~\ref{table} and \S~5.2), has a stellar mass
$1.6\times 10^{11}$~\msol\ and has an X--ray luminosity
$\simeq 3\times 10^{42}$ erg s$^{-1}$ \citep{pierre11}.  Neither the
near-IR spectrum \citep{tanaka10} nor IRAC colors of this galaxy show
any indication of an AGN. Nevertheless, if this X-ray flux stems from
accretion onto a SMBH, then it is consistent with the model
predictions.

There are also effects associated with the higher density region of
the cluster that could influence the galaxies'  morphological
evolution.     Interactions between the galaxies and the intracluster
medium (ICM) play a significant role in massive clusters
\citep[\eg,][]{mccarthy08,balogh09}.  However, it seems doubtful that
these influence galaxy evolution in this $z=1.62$ cluster, as
observations of its X-ray emission show that the hot ICM gas has not
developed fully \citep{pierre11}.  Therefore, effects associated with
interactions with the ICM are likely less important drivers of galaxy
evolution in this cluster \citep[see further discussion
in][]{mcgee09}.     

Galaxies associated with the cluster likely experience an
accelerated merger rate for the reason that this cluster is still
forming and has a high density of galaxies.   Galaxy assembly via
mergers is most effective in small groups and forming clusters at
lower redshifts, because these systems have lower velocity dispersions
\citep[see,][]{tran08,mcintosh08,mcgee09,wilman09}.  For example,
\citet{mcintosh08} find evidence that near--equal mass (``major'')
mergers between  red galaxies are more common in $z < 1$ groups than
in massive clusters.    It follows that mergers are also an important
assembly mechanism in forming clusters at higher redshift, and that
this process is accelerated in the higher density regions. 

Mergers are expected to be an important assembly mechanism for
massive, morphologically early-type galaxies.  Models show that at
late times ($z \lsim 2$)  these galaxies grow primarily through
dissipationless minor mergers and through the steady accretion of
smaller stellar systems formed outside the galaxies' virial radii
\citep{loeb03,ciotti07,naab09,oser10}.  These events increase the
galaxies' effective radii with a relatively mild increase in stellar
mass.  Measurements of the galaxy merger rate in high density regions
(such as clusters) show that these mergers occur mainly without star
formation \citep{elli10}. \citep[Mergers involving even small amounts
of  star formation are disfavored by the measured evolution of the
colors of cluster galaxies down to lower redshift,][]{vandokkum07a}.
A higher incidence of these dry mergers is expected to play a dominant
role in the evolution of quiescent galaxies at $z\lsim 2$ \citep[see,
\eg, discussion in][]{vanderwel11}.  

Our observation that quiescent galaxies in  the $z=1.62$ cluster have
larger sizes could be related to an accelerated dry merger rate
associated with the forming cluster.    There is some evidence to
support this as the massive quiescent galaxies in the $z=1.62$ cluster
appear to show a higher frequency of companions than those in the
field, which implies a higher current rate of mass growth from merging
\citep{lotz11}.  This is consistent with the models of \citet{shan12},
which predict that at fixed stellar mass central galaxies in larger
mass halos have large effective sizes compared to central galaxies in
lower-mass haloes.  In the Shankar et al.\ (2012) models, galaxies in
different halos masses undergo different types and numbers of mergers,
consistent with the results here.
%

While the observations suggest that cluster galaxies have experienced
an accelerated history at redshifts greater than $z=1.6$, there is
evidence that additional evolution is also required.  First, the most
massive galaxies in the cluster are still only $\sim$10-50\% as
massive as the brightest galaxies in low redshift clusters
\citep{blak06,holden09,vale10}.  These galaxies need to increase both
their stellar masses and their effective sizes by at least a factor of
2.      Simulations predict that the growth of massive cluster
galaxies at ``late'' times ($z < 1.5$) should occur more through the
dissipationless mergers of relatively massive progenitors
\citep[\eg,][]{delucia07a,rusz09}.  \citet{rusz09} predict that dry
major mergers are an important growth mechanism for galaxies at $z <
1.5$, and that the number of major mergers declines strongly with
galaxy mass.  If this is the case, then we may expect that the more
massive quiescent galaxies ($\gsim 10^{11}$~\msol) associated with
this cluster will experience $\sim$1--4 additional major dissipationless mergers (see
also \S~6.2).    This is similar to the findings of
\citet{vandokkum99} that 50\% of massive early-type massive galaxies
in the $z=0.83$ cluster MS 1054+03 will undergo a dry major merger at
$z < 1$.  

Second, the ellipticity distributions of the quiescent galaxies in the
$z=1.62$ cluster are shifted to relative high values, and these
galaxies show evidence for extended disks.  Both facts contrast
strongly with observations of early-type cluster galaxies at $z < 1$
(see \S~5).  Mergers would account both for the required evolution in
size, ellipticity, and mass as they transform the surface brightness
profiles toward higher \sersic\ indices \citep[see, \eg,][]{nava90}.

Lastly, measurements of the luminosity function of red-sequence
cluster galaxies show that the bright (massive) end is consistent with
passive evolution since $z \lsim 1$ \citep{rudn09}.  However, a
preliminary analysis shows that in the $z=1.62$ cluster here the
\textit{bright} end of the red-sequence-galaxy luminosity function is
not fully formed, and the massive galaxies require additional mass
growth mostly through dry mergers without substantial star formation
\cite{rudn12}.

Therefore, we conclude that the quiescent galaxies associated with the
$z=1.62$ cluster require additional growth through dry mergers to
match the properties of early-type massive galaxies in lower redshift
clusters.   This is consistent with the findings of  \citet{mcgrath08}
and \citet{vanderwel11} for  quiescent field galaxies at $z\sim 2$.
However, as discussed in \S~4.2, the quiescent galaxies in the
$z=1.62$ cluster appear to require \textit{less} size growth from
$z\sim 1.6$ to $z\sim 0$ compared to field galaxies in order to them
to be consistent with the  size-mass relation for quiescent galaxies
in the field and clusters \citep{wein09,vale10}.   Because  dissipationless
mergers of low-mass companions (``minor'' mergers) produce more size
growth relative to stellar-mass growth.  If quiescent galaxies grow
primarily through this mechanism, then it seems to follow that the
quiescent galaxies in the $z=1.62$ cluster will experience additional
mergers weighted toward more massive progenitors (i.e., they will
experience more major mergers), compared to quiescent galaxies in the
lower density field. 

\subsection{The formation of the brightest cluster galaxy?}

One interesting possibility is that the most massive  quiescent
galaxies in the $z=1.62$ cluster could merge into the brightest
cluster galaxy (BCG).  \citet{vale10} show that while early--type
galaxies in local clusters follow the size--mass relation of other
(field) early--type galaxies \citep[see also][]{wein09}, BCGs are
often outliers, having significantly larger effective radii for their
stellar mass \citep{bern07,rusz09}, although see \citet{lauer07} and
\citet{guo09} for alternative interpretations.   The most massive
galaxies in the $z=1.62$ cluster have stellar masses $\approx 2\times
10^{11}~\msol$, and these would require at least a factor of 2 growth
(and as much as a factor of 5) to achieve the stellar mass of the BCGs
in lower redshift clusters measured by \citet{vale10}.  

The two most massive galaxies in the  $z=1.62$ cluster, ID 39716 and
40170 (see figure~\ref{fig:sbfit}), are both near the core of the
cluster (each within a physical distance of $<$70 kpc), and they have
a projected physical separation of 126 kpc.   Assuming these galaxies
have relative velocities of $\gsim 100$ km s$^{-1}$ (about one third
the estimated velocity dispersion), they would experience a first pass
encounter in $\lsim 1$ Gyr.  It therefore seems entirely likely that
these galaxies will merge by $z \sim 1.2$.  This is consistent with
simulations that predict the progenitor of the BCG should experience
$1-2$ major mergers between $z\sim 1.5$ and $z \sim 1$
\citep[\eg,][]{delucia07a,rusz09}.   These galaxies currently have
stellar masses $1.5 \times 10^{11}$~\msol\ and $2.1 \times
10^{11}$~\msol, and effective radii, 2.6 and 3.2 kpc, respectively.
Assuming they will merge with no additional star formation, the
remnant will have a stellar mass $M >3 \times 10^{11}$~\msol, with a
more compact morphology (\sersic\ index $n \sim 4$), and grow in
effective radius to $\gsim$6 kpc, based on arguments from the virial
theorem \citep[see][]{nipoti03}.   Additional growth through mergers
and accretion (including accretion through dynamical friction of other
galaxies in the cluster potential) would increase the size and mass
further, shifting the new galaxy along (or even \textit{above}) the
size--mass relationship in figure~\ref{fig:sizemass} (consistent with
some low redshift BCGs, see Valentinuzzi et al.\ 2010).  Therefore, it
seems we are witnessing the progenitors of the BCG in this cluster
before they merge. 

\section{Summary}

In this paper we discussed morphological properties of galaxies in a
$z=1.62$ cluster \fulltarget\ using partial near-IR coverage from
\hst/WFC3 as part of CANDELS.   The cluster shows a prominent red
sequence dominated by galaxies with colors consistent with passive
evolution \citep{papo10a}, although there is a population of
star-forming galaxies in this cluster with high SFRs \citep{tran10}.
Recent \chandra\ data for this cluster show that the X-ray emission is
mostly attributed to point sources, suggesting that this cluster is
still in the process of collapse \citep{pierre11}.  Therefore, we are
able to study galaxy evolution in the high density region of a forming
cluster at high redshift. 

The \hst/WFC3 images show that the cluster galaxies exhibit a clear
color--morphology relation, where galaxies with colors of quiescent
stellar populations  have dominant spheroids, and galaxies with colors
consistent with ongoing star formation have disk--like and irregular
morphologies.  

The quiescent cluster galaxies follow a size--mass
relationship, but the cluster is deficient in quiescent galaxies with
compact effective radii compared to quiescent galaxies in the field at
$z=1.6$.    The average effective radii of the quiescent galaxies in
the cluster are larger compared to quiescent galaxies in the field at
fixed stellar mass ($\gsim 90$\% significance).  

If the difference in effective radii between the cluster and field
galaxies is generalizable, then it implies that the quiescent cluster galaxies experience
accelerated size evolution at redshifts greater than 1.6 compared to
similarly selected field galaxies.  Because other mechanisms
associated with interactions between the galaxies and the cluster ICM
are not yet operating, we argue that to explain the observations
quiescent cluster galaxies have had accelerated spheroid formation,
possibly as a result of the migration of stars formed in disks, and/or
merger histories associated with the formation of this cluster.  This
gives rise to an accelerated size growth compared to galaxies in the
field.  

The morphologies of quiescent galaxies in the field and cluster are
dominated by spheroids.  However, their ellipticity distributions are
broad, with median values $\epsilon_\mathrm{med} = 0.4$, with no trend
between ellipticity and mass, in contrast to lower redshift samples.
Both the ellipticity distributions and the surface-brightness profiles
of the massive cluster galaxies suggest these galaxies host stellar
disk components.    Because the quiescent galaxies have low estimated
SFRs, these disks are likely composed of stellar material with low gas
fractions available for star formation, either because they have
depleted their gas supplies, or that the dominant spheroids stabilize
the gas in the disks, hindering the formation of instabilities.   This
is true even for the massive quiescent cluster galaxies ($M>1\times
10^{11}$~\msol), which show no indications of star formation, are
spheroid--dominated, yet show extended emission consistent with disks
of scale lengths, 2--5~kpc.  The implication is that these galaxies
have significant stellar disks, similar to the interpretation of data
for passive galaxies in the field at $z\sim 1.5-2$
\citep{mcgrath08,vanderwel11}.  These extended disks are not present
in quiescent galaxies in clusters at lower redshifts ($z < 1$)
\citep{holden09}.  Therefore, some mechanism must transform or destroy
these disks in the few billion years from $z\sim 1.6$ to $z\sim 1$. 

The quiescent galaxies in the cluster at $z=1.62$ require additional
growth to match the observed properties of massive galaxies in
clusters at lower redshift.    Several lines of evidence suggest this
additional growth occurs via dissipationless (dry) mergers.  These
mergers will increase the sizes and stellar masses of the quiescent
galaxies, and affect the morphological transformation to more compact
surface--brightness profiles ($n \sim 4$).   However, because the
quiescent galaxies associated with the cluster at $z=1.62$ have larger
sizes, they appear to require \textit{slower} size growth at later
times ($z \lsim 1.6$) compared to galaxies in the field.    The
evidence for this is a result of comparing the data here with  results
from the literature.  The  size evolution of massive cluster galaxies
is relatively slow from $z\simeq 1.6$ to the present, with sizes
evolving as $(1+z)^{-0.6\pm0.1}$ compared to $\approx (1+z)^{-1}$ for
field galaxies \citep{vanderwel08}.  

To summarize, the data provide evidence that the morphology and size evolution in
the quiescent cluster galaxies  is accelerated compared to field
galaxies prior to  $z=1.6$ to account for the larger average sizes of
the quiescent cluster galaxies at this redshift.  In addition, we
conclude that additional growth is necessary for these galaxies and
that most of the growth occurs via dry mergers without significant
star formation.  Furthermore, in the case of the quiescent cluster
galaxies at $z=1.62$, much of this merger activity must occur between
$1 < z < 1.6$ such that these galaxies have the morphological
properties (\sersic\ indices and ellipticities) of cluster galaxies at
lower redshift.  This merger scenario appears consistent with
semianalytic model predictions, which predict that dissipationless
mergers dominate the mass growth of massive galaxies at $z \lsim 1.5$
\citep[\eg,][]{delucia07a,rusz09}. 

One caveat to these conclusions is that our analysis is based
on only a single cluster at $z=1.62$, which currently has
\hst/WFC3 imaging for $\sim 20$ quiescent cluster galaxies.   Clearly,
extending this analysis to galaxies in other clusters at $z > 1.5$ is
required to determine if the results here are generalized to
quiescent galaxies in other high density environments.

\acknowledgments

 We wish to thank the members of the CANDELS team for their
contributions to the dataset presented here.   We acknowledge
J. Blakeslee, M. Cooper, B. Holden, L.~Macri, R. Overzier, R. Quadri,
F. Shankar, J.~Stott, and S. Weinmann for helpful discussions and
comments, and we thank the anonymous referee for suggestions that
improved the presentation of the paper.  This work is based on
observations taken by the CANDELS Multi-Cycle Treasury Program with
the NASA/ESA HST, which is operated by the Association of Universities
for Research in Astronomy, Inc., under NASA contract NAS5-26555.  This
work is supported by HST program number GO-12060.  Support for Program
number GO-12060 was provided by NASA through a grant from the Space
Telescope Science Institute, which is operated by the Association of
Universities for Research in Astronomy, Incorporated, under NASA
contract NAS5-26555.  JSD acknowledges the support of the European
Research Council through an Advanced Grant, and the support of the
Royal Society via a Wolfson Research Merit Award.   This work is based
on observations made with the \textit{Spitzer Space Telescope}, which
is operated by the Jet Propulsion Laboratory, California Institute of
Technology.  This work is based in part on data obtained as part of
the UKIRT Infrared Deep Sky Survey.  We acknowledge generous support
from the Texas A\&M University and the George P.\ and Cynthia Woods
Institute for Fundamental Physics and Astronomy.

\bibliography{apj-jour,alpharefs}{}


\begin{deluxetable}{lccccccccccccc}
\ifsubmode
\rotate
\fi
\tablecaption{Properties of $z=1.6$ Galaxy Samples in the UDS CANDELS field\label{table}}
\tablecolumns{15}
\ifsubmode
      \tabletypesize{\footnotesize}
\else
\fi
\tablewidth{0pc}
\tablehead{
\colhead{ID} & 
\colhead{R.A.} & 
\colhead{Decl.} & 
\colhead{$z_\mathrm{ph}$} & 
\colhead{$\mathcal{P}_z$} & 
\colhead{$z - J$} & 
\colhead{$J - \mone$} & 
\colhead{$J_{125}^\mathrm{GALFIT}$} & 
\colhead{$R_\mathrm{eff}$} & 
\colhead{$n$} & 
\colhead{$\epsilon$} & 
\colhead{$\log M_\ast / \msol$} & \colhead{$\log \Psi / \msol$
  yr$^{-1}$} & 
\colhead{$R_\mathrm{proj}$} \\
\colhead{} & 
\colhead{(deg)} & 
\colhead{(deg)} & 
\colhead{} & 
\colhead{} & 
\colhead{(mag)} & 
\colhead{(mag)} & 
\colhead{(mag)} & 
\colhead{(kpc)} & 
\colhead{} & 
\colhead{} & 
\colhead{} & 
\colhead{} & 
\colhead{(Mpc)} \\
\colhead{(1)} & \colhead{(2)} & \colhead{(3)} & \colhead{(4)} & \colhead{(5)} & \colhead{(6)} & \colhead{(7)} & 
\colhead{(8)} & \colhead{(9)} & \colhead{(10)} & \colhead{(11)} &
\colhead{(12)} & \colhead{(13)} & \colhead{(14)}
}
\startdata
 39681  &      34.58987  & $-5.17487$  &  1.78  &  0.38  &  $1.40$  &  $2.84$  &  23.85  &  $ 3.4 \pm  1.5$ & $ 0.8 \pm  0.2$  &   0.61  &    \nodata &   \nodata &   0.04  \\
 39716  &      34.58789  & $-5.17585$  &  1.62  &  0.88  &  $1.64$  &  $2.00$  &  20.96  &  $ 2.6 \pm  0.1$ & $ 2.1 \pm  0.1$  &   0.45  &  $11.20^{+0.08}_{-0.04}$  &  $1.07$ &   0.06  \\
 40170  &      34.58979  & $-5.17218$  &  1.56  &  0.77  &  $1.52$  &  $2.14$  &  21.19  &  $ 3.2 \pm  0.2$ & $ 4.0 \pm  0.2$  &   0.29  &  $11.31^{+0.15}_{-0.08}$  &  $0.93$ &   0.07  \\
 39988  &      34.58759  & $-5.17225$  &  1.69  &  0.69  &  $1.64$  &  $1.57$  &  22.88  &  $ 0.9 \pm  0.1$ & $ 3.6 \pm  0.6$  &   0.22  &  $10.29^{+0.12}_{-0.10}$  &  $1.01$ &   0.07  \\
 39513  &      34.58626  & $-5.17594$  &  1.64  &  0.55  &  $1.93$  &  $1.78$  &  23.92  &  $ 1.0 \pm  0.2$ & $ 2.8 \pm  1.9$  &   0.14  &  $10.12^{+0.20}_{-0.28}$  &  $ < -1.0$ &   0.09  \\
 39770  &      34.59290  & $-5.17407$  &  1.67  &  0.51  &  $0.75$  &  $1.11$  &  22.95  &  $ 0.9 \pm  0.1$ & $ 2.9 \pm  0.4$  &   0.15  &  $ 9.68^{+0.19}_{-0.40}$  &  $1.73$ &   0.13  \\
 39462  &      34.59291  & $-5.17630$  &  1.64  &  0.50  &  $0.61$  &  $0.73$  &  23.11  &  $ 1.7 \pm  0.1$ & $ 1.0 \pm  0.1$  &   0.70  &  $ 9.27^{+0.32}_{-0.26}$  &  $1.18$ &   0.14  \\
 39218  &      34.58884  & $-5.17889$  &  1.65  &  0.86  &  $1.57$  &  $3.13$  &  22.83  &  $ 5.7 \pm  1.6$ & $ 0.4 \pm  0.1$  &   0.52  &  $11.16^{+0.22}_{-0.29}$  &  $1.27$ &   0.14  \\
 40387  &      34.59092  & $-5.16989$  &  1.73  &  0.52  &  $1.53$  &  $3.06$  &  22.32  &  $ 7.0 \pm  0.4$ & $ 1.5 \pm  0.1$  &   0.46  &  $11.09^{+0.20}_{-0.13}$  &  $0.17$ &   0.15  \\
 40249  &      34.58516  & $-5.17090$  &  1.58  &  0.74  &  $1.16$  &  $2.26$  &  22.19  &  $ 1.7 \pm  0.1$ & $ 3.3 \pm  0.2$  &   0.18  &  $10.53^{+0.23}_{-0.26}$  &  $1.94$ &   0.15  \\
 39062  &      34.59174  & $-5.17974$  &  1.68  &  0.53  &  $1.61$  &  $2.19$  &  23.47  &  $ 1.0 \pm  0.1$ & $ 2.3 \pm  0.6$  &   0.14  &  $10.53^{+0.19}_{-0.12}$  &  $-0.31$ &   0.19  \\
 40299  &      34.59352  & $-5.16947$  &  1.58  &  0.45  &  $0.80$  &  $1.69$  &  35.36  &  $ 2.9 \pm  7.2$ & $ 0.4 \pm  1.0$  &   0.36  &  $ 9.99^{+0.23}_{-0.37}$  &  $1.28$ &   0.20  \\
 40449  &      34.58736  & $-5.16763$  &  1.69  &  0.44  &  $0.88$  &  $0.73$  &  23.95  &  $ 2.7 \pm  1.8$ & $ 1.5 \pm  0.9$  &   0.48  &  $ 9.00^{+0.40}_{-0.22}$  &  $0.95$ &   0.21  \\
 39858  &      34.58199  & $-5.17316$  &  1.60  &  0.54  &  $0.89$  &  $1.48$  &  22.90  &  $ 2.4 \pm  0.3$ & $ 0.5 \pm  0.1$  &   0.57  &  $ 9.82^{+0.22}_{-0.40}$  &  $1.53$ &   0.21  \\
 39230  &      34.59285  & $-5.17985$  &  1.63  &  0.69  &  $1.50$  &  $2.44$  &  22.27  &  $ 1.7 \pm  0.4$ & $ 4.0 \pm  0.5$  &   0.63  &  $11.04^{+0.15}_{-0.15}$  &  $1.55$ &   0.21  \\
 40422  &      34.58407  & $-5.16859$  &  1.84  &  0.36  &  $1.31$  &  $1.56$  &  23.59  &  $ 3.3 \pm  0.9$ & $ 1.0 \pm  0.3$  &   0.83  &    \nodata &   \nodata &   0.22  \\
 39395  &      34.58038  & $-5.17745$  &  1.67  &  0.77  &  $1.82$  &  $2.30$  &  22.48  &  $ 1.2 \pm  0.1$ & $ 3.2 \pm  0.5$  &   0.17  &  $10.76^{+0.11}_{-0.05}$  &  $ < -1.0$ &   0.27  \\
 40238  &      34.58054  & $-5.17040$  &  1.73  &  0.50  &  $1.80$  &  $1.98$  &  22.27  &  $ 1.0 \pm  0.1$ & $ 1.5 \pm  0.1$  &   0.78  &  $10.81^{+0.04}_{-0.04}$  &  $0.46$ &   0.28  \\
 40728  &      34.58487  & $-5.16563$  &  1.77  &  0.42  &  $0.78$  &  $0.71$  &  23.42  &  $ 2.3 \pm  0.3$ & $ 1.1 \pm  0.2$  &   0.40  &  $ 9.32^{+0.24}_{-0.37}$  &  $0.83$ &   0.29  \\
 40164  &      34.57984  & $-5.17074$  &  1.68  &  0.47  &  $0.74$  &    \nodata & 22.69  &  $ 3.3 \pm  0.2$ & $ 0.5 \pm  0.1$  &   0.40  &  $ 9.99^{+0.16}_{-0.19}$  &  $1.68$ &   0.29  \\
 40730  &      34.59290  & $-5.16535$  &  1.74  &  0.43  &  $0.60$  &  $0.46$  &  23.86  &  $ 1.7 \pm  0.2$ & $ 0.4 \pm  0.1$  &   0.52  &  $ 9.00^{+0.34}_{-0.31}$  &  $0.86$ &   0.30  \\
 40748  &      34.58281  & $-5.16616$  &  1.65  &  0.88  &  $1.57$  &  $1.45$  &  21.96  &  $ 1.8 \pm  0.2$ & $ 5.1 \pm  0.4$  &   0.27  &  $10.60^{+0.03}_{-0.03}$  &  $ < -1.0$ &   0.31  \\
 39175  &      34.57958  & $-5.17840$  &  1.69  &  0.53  &  $0.80$  &    \nodata & 23.21  &  $ 3.5 \pm  0.7$ & $ 2.4 \pm  0.5$  &   0.25  &  $ 9.49^{+0.23}_{-0.31}$  &  $ < -1.0$ &   0.31  \\
 40567  &      34.58071  & $-5.16696$  &  1.77  &  0.40  &  $0.63$  &  $1.02$  &  23.50  &  $ 0.9 \pm  0.1$ & $ 2.6 \pm  0.8$  &   0.62  &  $ 9.06^{+0.38}_{-0.17}$  &  $1.12$ &   0.33  \\
 40456  &      34.57959  & $-5.16800$  &  1.58  &  0.49  &  $1.03$  &  $1.38$  &  23.49  &  $ 2.0 \pm  0.2$ & $ 0.5 \pm  0.1$  &   0.36  &  $ 9.63^{+0.25}_{-0.39}$  &  $1.35$ &   0.34  \\
 39600  &      34.57702  & $-5.17539$  &  1.63  &  0.63  &  $0.93$  &  $1.46$  &  23.19  &  $ 2.3 \pm  0.2$ & $ 0.6 \pm  0.1$  &   0.47  &  $ 9.59^{+0.33}_{-0.29}$  &  $1.50$ &   0.36  \\
 38665  &      34.57904  & $-5.18394$  &  1.65  &  0.59  &  $0.97$  &  $1.50$  &  22.74  &  $ 2.7 \pm  0.1$ & $ 0.7 \pm  0.1$  &   0.49  &  $10.12^{+0.18}_{-0.19}$  &  $1.54$ &   0.42  \\
 38030  &      34.58667  & $-5.18881$  &  1.49  &  0.38  &  $0.87$  &  $2.58$  &  22.95  &  $ 3.7 \pm  1.1$ & $ 0.6 \pm  0.1$  &   0.80  &    \nodata &   \nodata &   0.45  \\
 40606  &      34.57457  & $-5.16725$  &  1.55  &  0.58  &  $1.47$  &  $2.32$  &  22.62  &  $ 1.6 \pm  0.2$ & $ 3.9 \pm  0.4$  &   0.31  &  $10.55^{+0.12}_{-0.10}$  &  $2.05$ &   0.48  \\
 40901  &      34.57618  & $-5.16421$  &  1.80  &  0.30  &  $0.94$  &  $1.94$  &  23.32  &  $ 2.0 \pm  0.2$ & $ 1.4 \pm  0.4$  &   0.50  &    \nodata &   \nodata &   0.49  \\
 41634  &      34.58741  & $-5.15799$  &  1.74  &  0.46  &  $0.87$  &  $1.26$  &  23.20  &  $ 1.3 \pm  0.1$ & $ 1.0 \pm  0.1$  &   0.51  &  $ 9.54^{+0.36}_{-0.21}$  &  $1.57$ &   0.50  \\
 40640  &      34.57339  & $-5.16783$  &  1.61  &  0.89  &  $1.75$  &  $1.93$  &  21.37  &  $ 1.2 \pm  0.1$ & $ 3.5 \pm  0.2$  &   0.66  &  $11.15^{+0.05}_{-0.03}$  &  $ < -1.0$ &   0.51  \\
 40749  &      34.57211  & $-5.16608$  &  1.60  &  0.79  &  $1.67$  &  $1.40$  &  23.35  &  $ 0.6 \pm  0.1$ & $ 3.0 \pm  1.1$  &   0.51  &  $10.10^{+0.05}_{-0.06}$  &  $-0.84$ &   0.56  \\
 40064  &      34.57000  & $-5.17146$  &  1.57  &  0.42  &  $0.96$  &  $2.36$  &  23.78  &  $ 2.7 \pm  1.0$ & $ 0.3 \pm  0.1$  &   0.58  &  $10.04^{+0.37}_{-0.61}$  &  $0.94$ &   0.58  \\
 41493  &      34.57659  & $-5.15967$  &  1.84  &  0.30  &  $1.22$  &  $2.04$  &  23.36  &  $ 2.5 \pm  0.3$ & $ 0.9 \pm  0.1$  &   0.30  &    \nodata &   \nodata &   0.58  \\
 41510  &      34.57689  & $-5.15889$  &  1.87  &  0.30  &  $1.63$  &  $1.93$  &  24.26  &  $ 1.8 \pm  0.7$ & $ 1.6 \pm  1.3$  &   0.04  &    \nodata &   \nodata &   0.59  \\
 41874  &      34.57900  & $-5.15692$  &  1.65  &  0.69  &  $1.70$  &  $2.32$  &  21.63  &  $ 5.8 \pm  0.6$ & $ 5.0 \pm  0.4$  &   0.08  &  $10.97^{+0.07}_{-0.03}$  &  $0.25$ &   0.61  \\
 40973  &      34.57127  & $-5.16463$  &  1.68  &  0.72  &  $1.70$  &  $2.14$  &  21.89  &  $ 2.6 \pm  0.1$ & $ 2.6 \pm  0.2$  &   0.05  &  $10.94^{+0.09}_{-0.04}$  &  $1.12$ &   0.61  \\
 38582  &      34.57168  & $-5.18492$  &  1.71  &  0.61  &  $1.57$  &  $1.92$  &  21.81  &  $ 2.0 \pm  0.1$ & $ 1.3 \pm  0.1$  &   0.36  &  $10.89^{+0.05}_{-0.05}$  &  $ < -1.0$ &   0.61  \\
 40928  &      34.57100  & $-5.16411$  &  1.68  &  0.48  &  $1.52$  &    \nodata & 24.12  &  $ 2.0 \pm  0.3$ & $ 0.7 \pm  0.3$  &   0.51  &  $10.16^{+0.28}_{-0.31}$  &  $1.49$ &   0.62  \\
 39097  &      34.56828  & $-5.17913$  &  1.65  &  0.41  &  $0.98$  &  $1.65$  &  24.34  &  $ 1.3 \pm  0.2$ & $ 0.7 \pm  0.1$  &   0.59  &  $ 9.23^{+0.48}_{-0.22}$  &  $1.06$ &   0.64  \\
 37794  &      34.57487  & $-5.19073$  &  1.85  &  0.31  &  $0.70$  &  $1.03$  &  23.41  &  $ 3.4 \pm  1.1$ & $ 1.9 \pm  0.8$  &   0.37  &    \nodata &   \nodata &   0.66  \\
 42331  &      34.59231  & $-5.15224$  &  1.72  &  0.53  &  $0.86$  &    \nodata & 22.90  &  $ 2.0 \pm  0.3$ & $ 2.0 \pm  0.3$  &   0.56  &  $ 9.96^{+0.10}_{-0.15}$  &  $1.30$ &   0.68  \\
 37856  &      34.57019  & $-5.19104$  &  1.81  &  0.35  &  $1.87$  &  $2.32$  &  23.19  &  $ 0.6 \pm  0.1$ & $ 3.1 \pm  0.5$  &   0.66  &    \nodata &   \nodata &   0.76  \\
 41989  &      34.57235  & $-5.15454$  &  1.69  &  0.50  &  $0.44$  &  $0.33$  &  23.15  &  $ 2.4 \pm  0.3$ & $ 0.6 \pm  0.1$  &   0.49  &  $ 8.88^{+0.33}_{-0.12}$  &  $ < -1.0$ &   0.78  \\
 38691  &      34.56440  & $-5.18374$  &  1.68  &  0.32  &  $1.14$  &  $2.58$  &  23.79  &  $ 2.1 \pm  0.8$ & $ 0.2 \pm  0.1$  &   0.75  &    \nodata &   \nodata &   0.79  \\
 42585  &      34.57853  & $-5.14985$  &  1.77  &  0.42  &  $1.51$  &  $2.08$  &  23.21  &  $ 1.2 \pm  0.3$ & $ 4.1 \pm  0.9$  &   0.36  &  $10.71^{+0.19}_{-0.13}$  &  $0.83$ &   0.81  \\
 42623  &      34.57957  & $-5.14849$  &  1.74  &  0.44  &  $0.88$  &  $1.05$  &  23.79  &  $ 1.2 \pm  0.2$ & $ 0.7 \pm  0.2$  &   0.73  &  $ 9.38^{+0.33}_{-0.35}$  &  $1.02$ &   0.83  \\
 42925  &      34.59246  & $-5.14639$  &  1.51  &  0.32  &  $1.37$  &  $2.06$  &  23.50  &  $ 1.6 \pm  0.5$ & $ 3.2 \pm  1.1$  &   0.37  &    \nodata &   \nodata &   0.86  \\
 37268  &      34.56874  & $-5.19596$  &  1.63  &  0.45  &  $0.98$  &  $1.65$  &  23.14  &  $ 2.5 \pm  0.4$ & $ 2.6 \pm  0.5$  &   0.50  &  $10.27^{+0.13}_{-0.20}$  &  $1.13$ &   0.90  \\
   \enddata
\tablecomments{A portion of the table is shown here. (1) Object ID,
  (2) right ascension (J2000), (3) declination (J2000), (4)
  photometric redshift, (5) integrated photometric redshift
  probability distribution function, see \S~2.2, (6) Suprime $z$, UKIDSS $J$
  color, (7) UKIDSS $J$, \spitzer/IRAC 3.6~\micron\ color, (8) WFC3
  F125 magnitude measured from GALFIT, (9) circularized effective
  radius, (10) \sersic\ index, (11) ellipticity, $\epsilon = 1 - b/a$,
  (12) stellar mass and (13) star formation rate from analysis of spectral energy distribution, (14) projected distance from center of the $z=1.62$ cluster \fulltarget.}
\end{deluxetable}

\end{document}